%% file: spiral.tex
\documentclass[aps,prb,twocolumn,floatfix,superscriptaddress,longbibliography]{revtex4-1}
\usepackage{graphicx}
\usepackage{bm}
\usepackage{color}
\usepackage{subfigure}
\usepackage[utf8]{inputenc}
\usepackage{amssymb}
\usepackage{hyperref}
\usepackage[normalem]{ulem}
\begin{document}
\renewcommand{\textfraction}{0.05} 

\title{Topological superconductivity at finite temperatures\\ 
in proximitized magnetic nanowires}
\author{Anna Gorczyca--Goraj}
\affiliation{Department of Theoretical Physics, University of Silesia, Katowice, Poland}
\author{Tadeusz Doma\'nski}
\affiliation{Institute of Physics, M. Curie Skłodowska University, Lublin, Poland}
\author{Maciej M. Ma\'ska}
\email{maciej.maska@phys.us.edu.pl}
\affiliation{Department of Theoretical Physics, University of Silesia, Katowice, Poland}

\date{\today}
\begin{abstract}

Performing Monte Carlo simulations we study the temperature dependent self--organization of magnetic moments coupled to itinerant electrons in a finite--size one--dimensional nanostructure proximitized to a superconducting reservoir. At low temperature an effective interaction between the localized magnetic moments, that is mediated by itinerant electrons, leads to their helical ordering. This ordering, in turn, affects the itinerant electrons, inducing the topologically nontrivial superconducting phase that hosts the Majorana modes. In a~wide range of system parameters, the spatial periodicity of a spiral order that minimizes the ground state energy turns out to promote the topological phase. We determine the correlation length of such spiral order and study how it is reduced by thermal fluctuations. This reduction is accompanied by suppression of the topological gap (which separates the zero-energy mode from continuum), setting the upper (critical) temperature for existence of the Majorana quasiparticles. Monte Carlo simulations do not rely on any ansatz for configurations of the localized moments, therefore they can be performed for arbitrary model parameters, also beyond the perturbative regime.

\end{abstract}
\maketitle

\section{Introduction}
Recent progress in fabricating artificial nanostructures with spatial constraints~\cite{santos.deen.15}  
enabled observation of novel quantum states~\cite{Zhang2018}, 
where topology plays a prominent role. Motivated by the seminal Kitaev's paper \cite{kitaev.01}, 
one of such intensively explored fields is related to topological 
superconductivity which occurs in semiconducting nanowires proximitized to superconductors~\cite{deng.yu.12,mourik.zuo.12,das.ronen.12,finck.vanharlingen.13,deng.vaitiekenas.16,nichele.drachmann.17,lutchyn.bakkers.18,gul.zhang.18} or nanochains of magnetic atoms  deposited on superconducting surfaces~\cite{nadjperge.drozdov.14,pawlak.kisiel.16,feldman.randeria.16,ruby.heinrich.17,jeon.xie.17,kim.palaciomorales.18}.
In both cases the Majorana-type quasiparticles have been observed at boundaries of 
proximitized nanoscopic wires/chains and non--Abelian statistics~\cite{nayak.simon.08} makes them promising  for realization of quantum computing~\cite{aasen.hell.16,karzig.knapp.17} and/or new spintronic devices \cite{Liu2016}.

Mechanism that drives the proximitized nanowire into a topologically non-trivial phase can originate 
from the spin-orbit coupling (SOC) combined with the Zeeman splitting above some critical value of 
magnetic field~\cite{sato.fujimoto.09,sato.takahashi.09,sato.takahashi.10,klinovaja.loss.12}. 
Upon approaching this transition a pair of finite-energy (Andreev) bound states 
coalesces into the degenerate Majorana quasiparticles~\cite{chevallier.sticlet.12,chevallier.simon.13} 
formed near the ends of nanowire. Another scenario combines the proximity--induced superconducting state with the spiral magnetic order 
\cite{choy.11,Martin.Morpurgo.12,Kjaergaard2012,Bernevig2013,Simon2013,Pientka2013,klinovaja.stano.13,Vazifeh.Franz.13,Franz2014,Li2014,Pientka2014,Kim2014,Heimes2014,Peng2015,Scalettar2015,Brydon2015,Schecter2015,heimes.mendler.15,Scalettar2015errata,Simon2015,Paaske2016}.
The latter approach is particularly appealing, because magnetic order seems to 
self-adjust its periodicity (characterized by the pitch vector $q_{*}$) to support 
the topological phase. Origin of the topological phase in a system with spirally ordered magnetic moments is mathematically equivalent to the scenario based on the spin--orbit and 
Zeeman interactions~{\cite{Braunecker2010,klinovaja.stano.13}} and its {\em topofilia} has been investigated by a number of groups~\cite{Vazifeh.Franz.13,Simon2013,Scalettar2015,Simon2015,Paaske2016}. 

Topological features of the  systems with self--organized spiral ordering have been so far studied, focusing mainly on the zero temperature limit. Thermal effects have been partly addressed, taking into account magnon excitations (which suppress a magnitude of the spiral order)~\cite{klinovaja.stano.13,Scalettar2015} and investigating a contribution of the entropy term to the free energy (which substantially affects the wave vector of the spiral order, so that magnetic order might be preserved but the electronic state could no longer be topological)~\cite{Simon2015}. Usually, however, any long--range order hardly exists in one--dimensional systems at finite temperatures and therefore it is important -- especially for practical applications of such systems -- to estimate the maximum temperature up to which the topologically nontrivial states could survive. For its reliable determination we perform here the  Monte Carlo (MC) simulations. 

Our numerical results unambiguously indicate that thermal effects are detrimental to both the topological superconducting state and to the Majorana quasiparticles. This is evidenced by: 
\begin{itemize}
\item[{\em (i)}] changeover of the topological ${\mathbb Z}_{2}$ number, 
\item[{\em (ii)}] characteristic scaling of the temperature-dependent coherence length of the spiral magnetic order, 
\item[{\em (iii)}] and directly from the quasiparticle spectrum, where thermal effects suppress the topological  energy gap converting the zero-energy quasiparticles into overdamped modes.  
\end{itemize}

The rest of the paper is organized as follows. In Sec.~\ref{sec:Model} we introduce the microscopic model. Next, in Sec.~\ref{sec:TZero}, we briefly revisit  the topologically nontrivial  superconducting state at zero temperature and check if it really coincides with the spiral pitch $q_{*}$ that minimizes the ground state energy. Essential results of our study are presented in Sec.\ \ref{sec:fin-temp}, where we analyze (dis)ordering of the magnetic moments at finite temperatures by means of the MC method  determining the upper (critical) temperature for existence of the topological superconducting state and the Majorana quasiparticles. Finally,  in Sec.~\ref{sec:concl}, we summarize the main results.

\section{Model}
\label{sec:Model}

We consider a chain of the localized magnetic impurities whose moments are coupled to the spins of itinerant electrons. This nanoscopic chain is deposited on a surface of  $s$--wave bulk superconductor,  through the proximity effect inducing electron pairing. Such system can be described by the following Hamiltonian
\begin{eqnarray}
    H=&-&t\sum_{i,\sigma}\hat{c}^\dagger_{i,\sigma}\hat{c}_{i+1,\sigma}
    -\mu\sum_{i,\sigma}\hat{c}^\dagger_{i,\sigma}\hat{c}_{i,\sigma}
    \label{eq:hamil}\\
    &+&J\sum_i{\bm S}_i\cdot\hat{\bm s}_i
    +\sum_i \left( \Delta \hat{c}^{\dagger}_{i\uparrow}\hat{c}^{\dagger}_{i\downarrow}
    +\mbox{\rm H.c.}\right),
    \nonumber
\end{eqnarray}
where $\hat{c}^\dagger_{i,\sigma}$ and $\hat{c}_{i,\sigma}$ are the creation and annihilation operators of electron at site $i$ and $\hat{\bm s}_i$ is their spin
\begin{equation}
\hat{\bm s}_i=\frac{1}{2}\sum_{\alpha,\beta}\hat{c}^\dagger_{i,\alpha}{\bm \sigma}_{\alpha\beta}
\hat{c}_{i,\beta}
\end{equation}
with ${\bm\sigma}$ being a vector of the Pauli matrices. We assume that magnetic moments ${\bm S}_i$ have much slower dynamics than electrons and can be treated classically. In general, they can be expressed in the spherical coordinates in terms of the polar and azimuthal angles $\theta_i$ and $\phi_i$
\begin{equation}
    {\bm S}_i = S\left(\sin\theta_i\cos\phi_i,\:\sin\theta_i\sin\phi_i,\:\cos\theta_i\right).
    \label{eq:parametrization}
\end{equation}
In the weak coupling $J$ limit it has been shown ~\cite{Simon2013,klinovaja.stano.13,Vazifeh.Franz.13} that the effective Ruderman-Kittel-Kasuya-Yosida interaction  induces the helical ordering between the magnetic moments of the impurities
\begin{equation}
    \phi_i = i \, a \, q_{*}
    \label{eq:pitch}
\end{equation}
where $a$ is the lattice constant and the spiral pitch $q_{*}$ is strongly dependent on the model parameters \cite{Scalettar2015,Scalettar2015errata}. Since Hamiltonian (\ref{eq:hamil}) has an SU(2) spin rotation symmetry, for any constant opening angle $\theta_i$ without loss of generality it can be assumed that $\theta_i=\pi/2$. It is possible to perform the gauge
transformation, upon which the localized magnetic moments become ferromagnetically
polarized at expense of introducing the spin and $q_{*}$--dependent hopping amplitude \cite{Martin.Morpurgo.12}. Here, however, we are mostly interested in nonzero temperatures,
where the ground state ordering is affected by thermal excitations. Therefore, we will treat
$\phi_i$'s as fluctuating degrees of freedom. This will allow us not only to describe thermal states, 
but also to take into account possible phase separation, where orderings with different values of $q_{*}$
take place in segments of the nanochain \cite{Scalettar2015,Scalettar2015errata}. We shall also check influence of $\theta_i$ fluctuations on stability of the topological phase (Sec.~\ref{sec:3Dspiral}).

In what follows we set the intersite spacing as a unit ($a=1$) and impose $S=1$\footnote{Formally we work in the large--spin regime $S\to\infty$ with a finite value of $JS=\mbox{const}$. For the sake of simplicity we assume $S=1$ and use $J$ to measure the coupling between the localized moments and the electron spin.}. For simplicity we also set the Boltzmann constant $k_{B}\equiv 1$ and treat the hopping integral as a convenient unit ($t=1$) for all energies discussed in our study.

\section{Topofilia of the ground state}
\label{sec:TZero}

\begin{figure}[b!]
\centering
\includegraphics[width=0.48\textwidth]{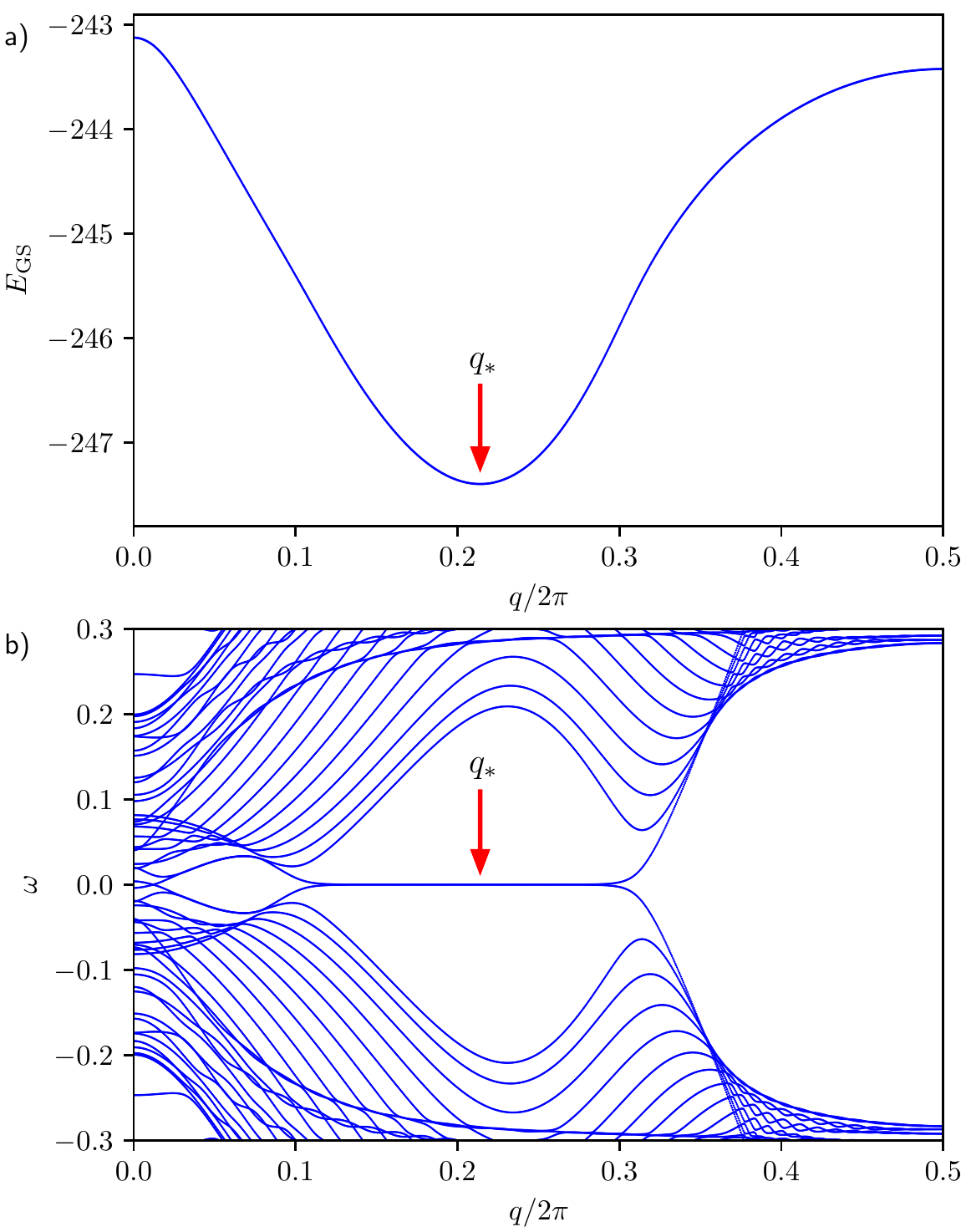}
\caption{a) The ground state energy $E_{\rm GS}$ versus the spiral pitch $q$ obtained for $\Delta=0.3,\: \mu=1.5$ and $J=1$. b) Evolution of the quasiparticle spectrum with respect to $q$. The red arrows indicate $q_{*}$, minimizing the ground state energy.}
\label{fig:gs_en}
\end{figure}

In the case of periodic boundary conditions a spin-dependent gauge transformation can convert the Hamiltonian (\ref{eq:hamil}) into a translationally invariant form that can be easily diagonalized\cite{Martin.Morpurgo.12,Vazifeh.Franz.13}. Here, however, we focus on the open boundary conditions what allows us to study the Majorana end states. Additionally, open boundary conditions do not impose any restrictions on the spiral pitch $q$, what is especially important for rather short nanochains.
Most of our calculations have been performed for the nanowire comprising 70 sites. We have numerically diagonalized the system, considering various configurations of the local magnetic moments $S_{i}$. In particular, we have inspected the spiral ordering 
and considered $q\in \left[ 0 ; \pi \right]$ varying the model parameters $J,\:\mu,\:\Delta$ (transformation $q\to -q$ changes the chirality of the spiral but it neither affects the thermodynamic nor topological properties).

\begin{figure*}
    \centering
    \includegraphics[width=0.44\textwidth]{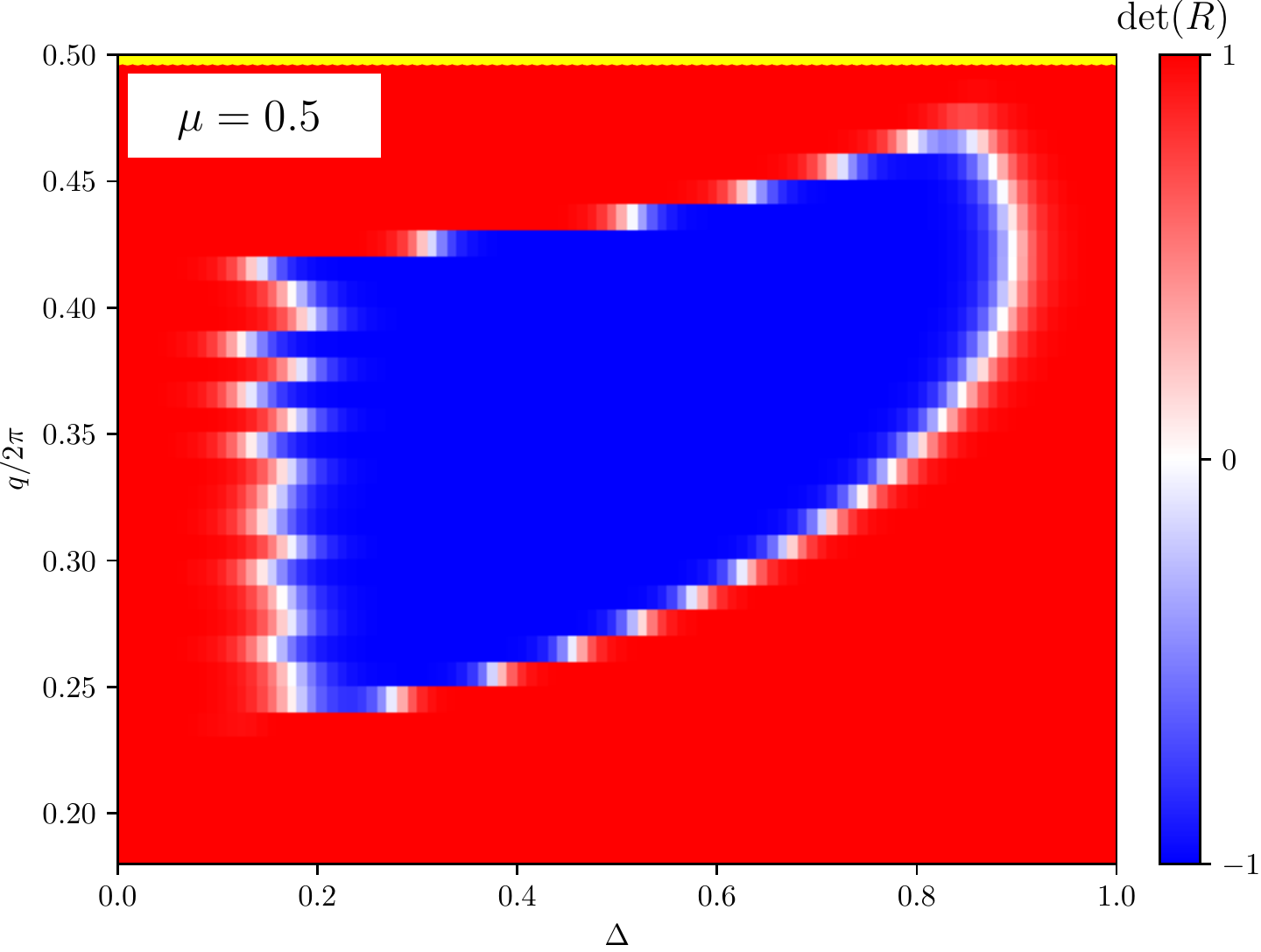}
    \includegraphics[width=0.44\textwidth]{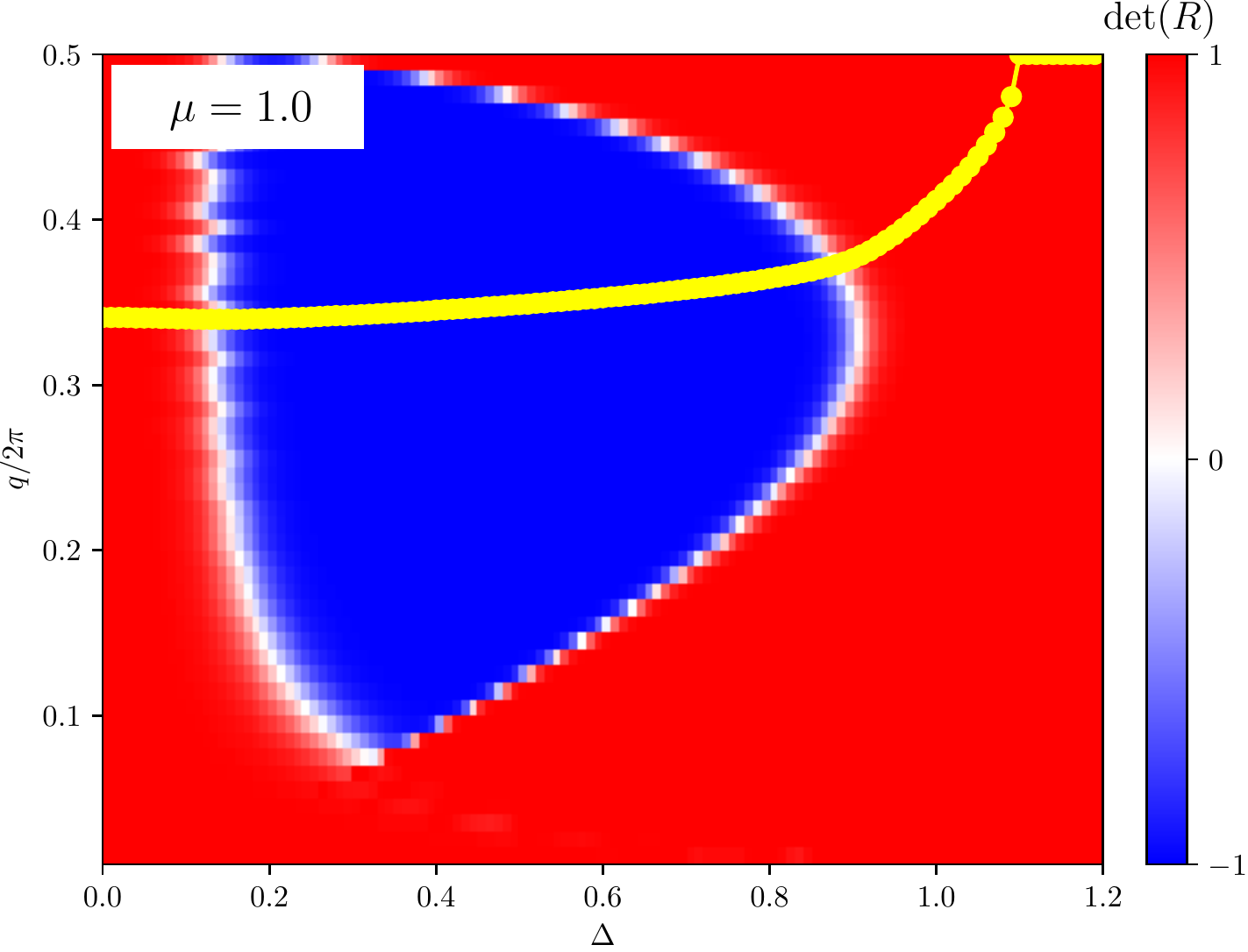}
    \includegraphics[width=0.44\textwidth]{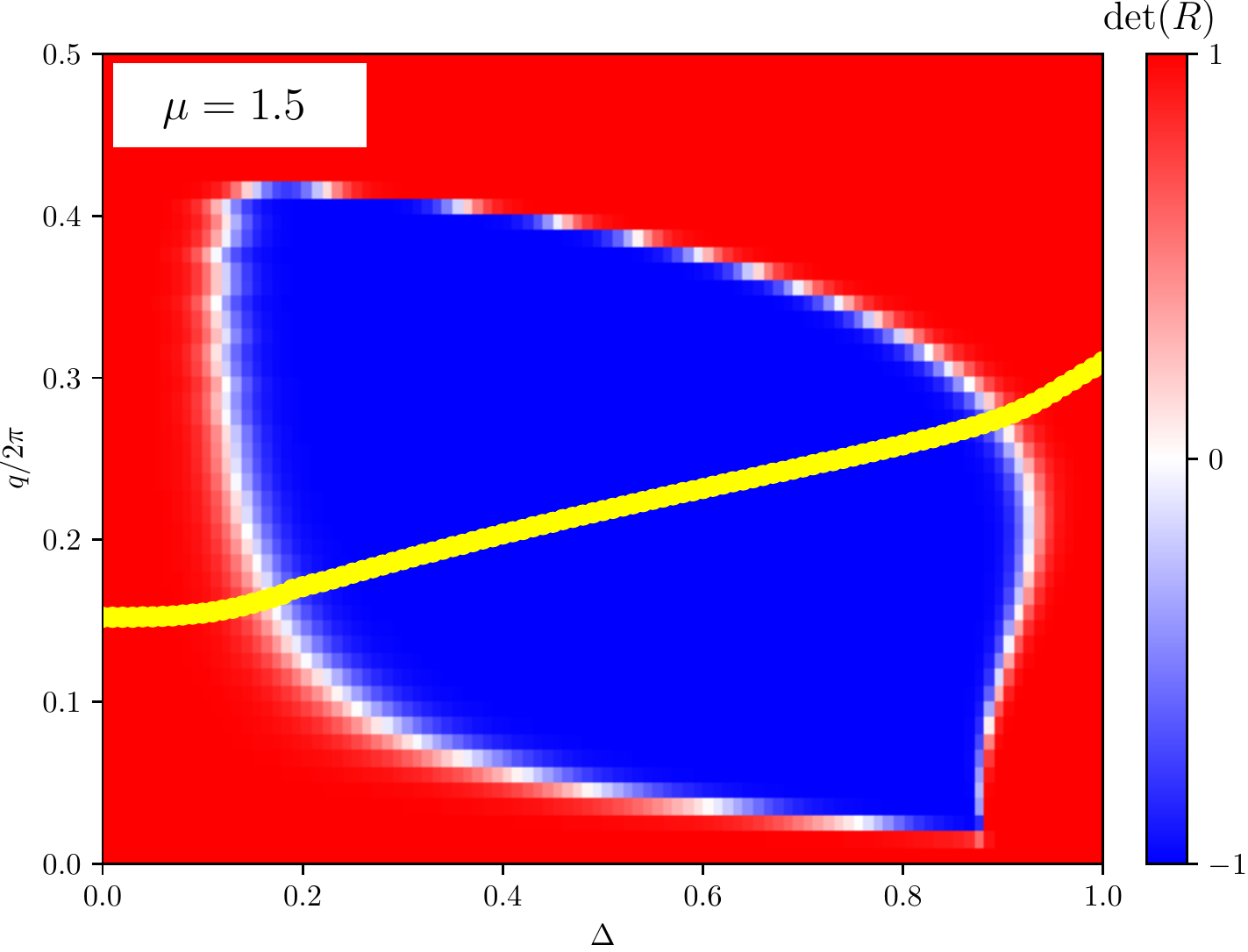}
    \includegraphics[width=0.44\textwidth]{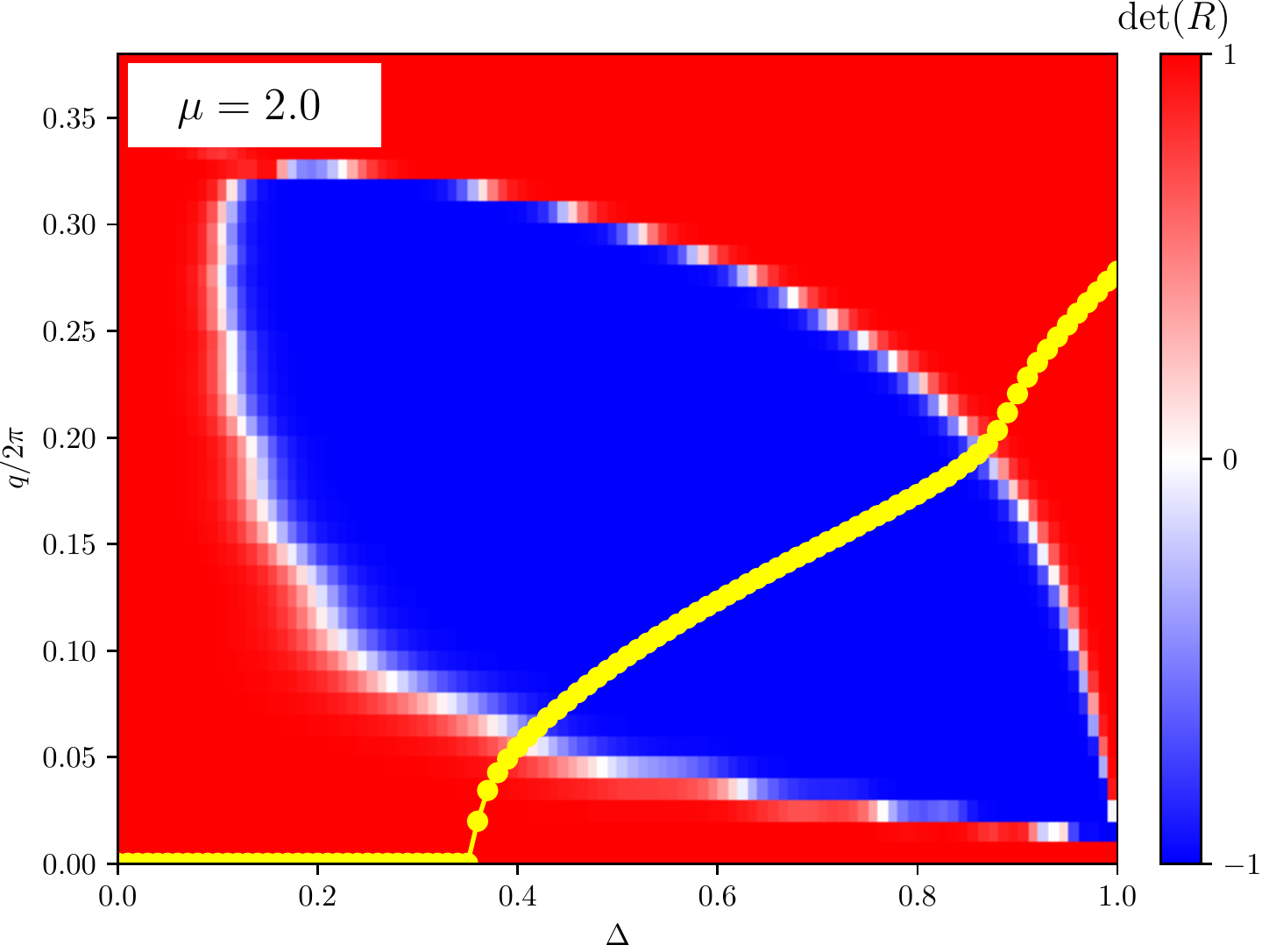}
    \includegraphics[width=0.44\textwidth]{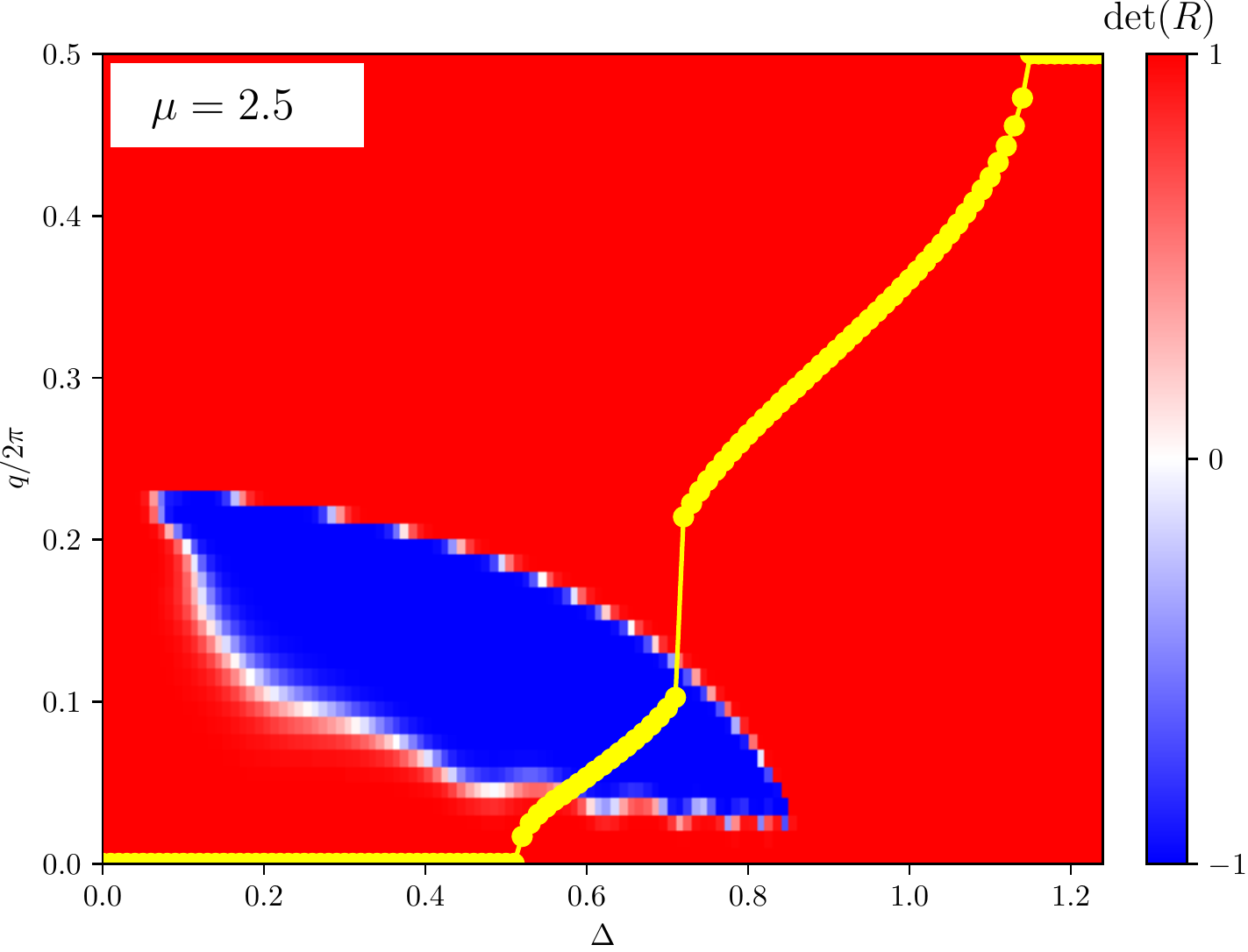}
    \includegraphics[width=0.44\textwidth]{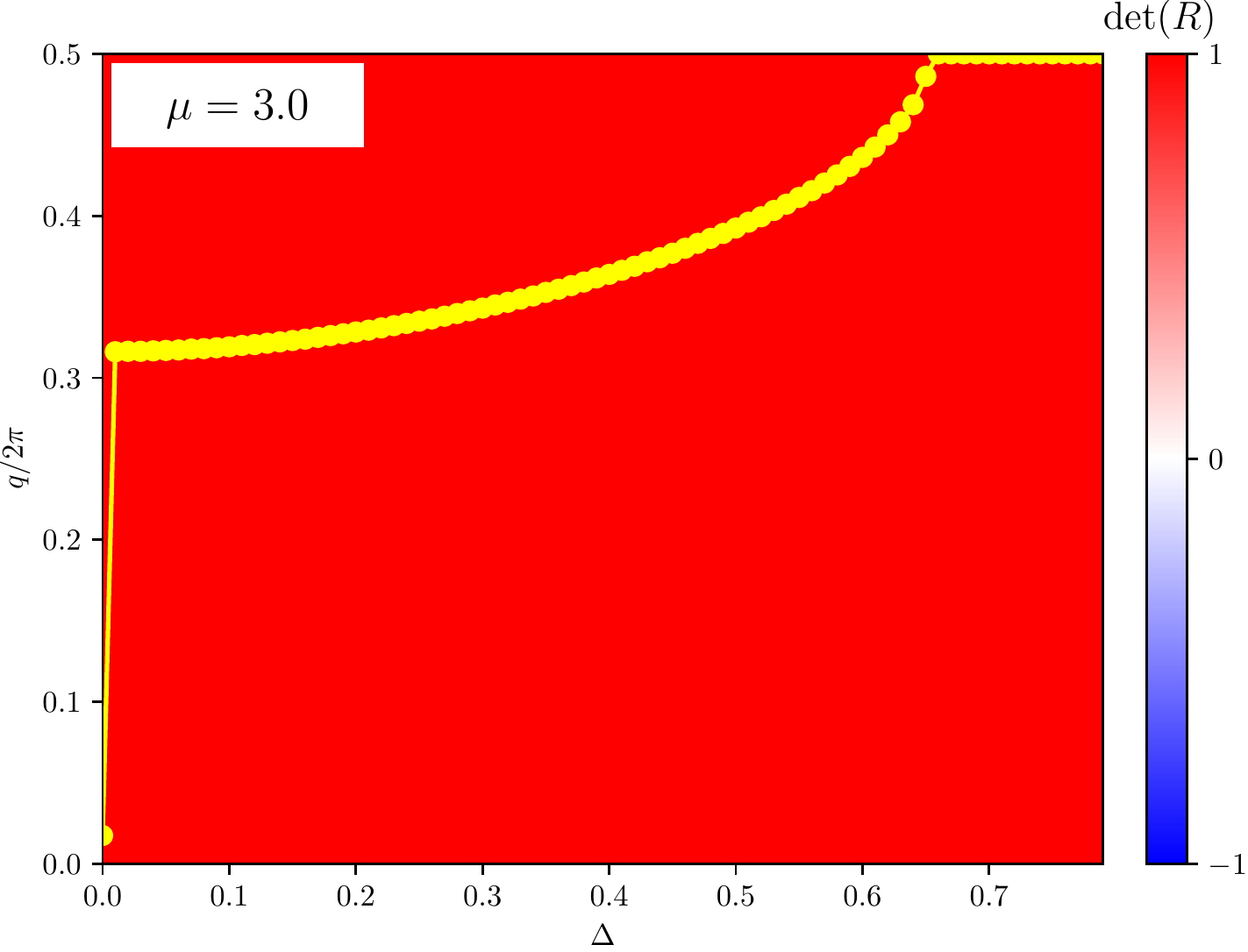}
    \caption{Zero temperature value of $\det(R)$ [see Eq. (\ref{topo_number})] as a function of $\Delta$ and $q$ obtained for 70 sites, using $J=2$ and different values of $\mu$. The yellow circles show $q_*$, minimizing the ground state energy. Note, that for $\mu=0.5$ $q_*$ is equal to 0.5.}
    \label{fig:topo_qmin_J=2}
\end{figure*}

\begin{figure*}
    \centering
    \includegraphics[width=0.44\textwidth]{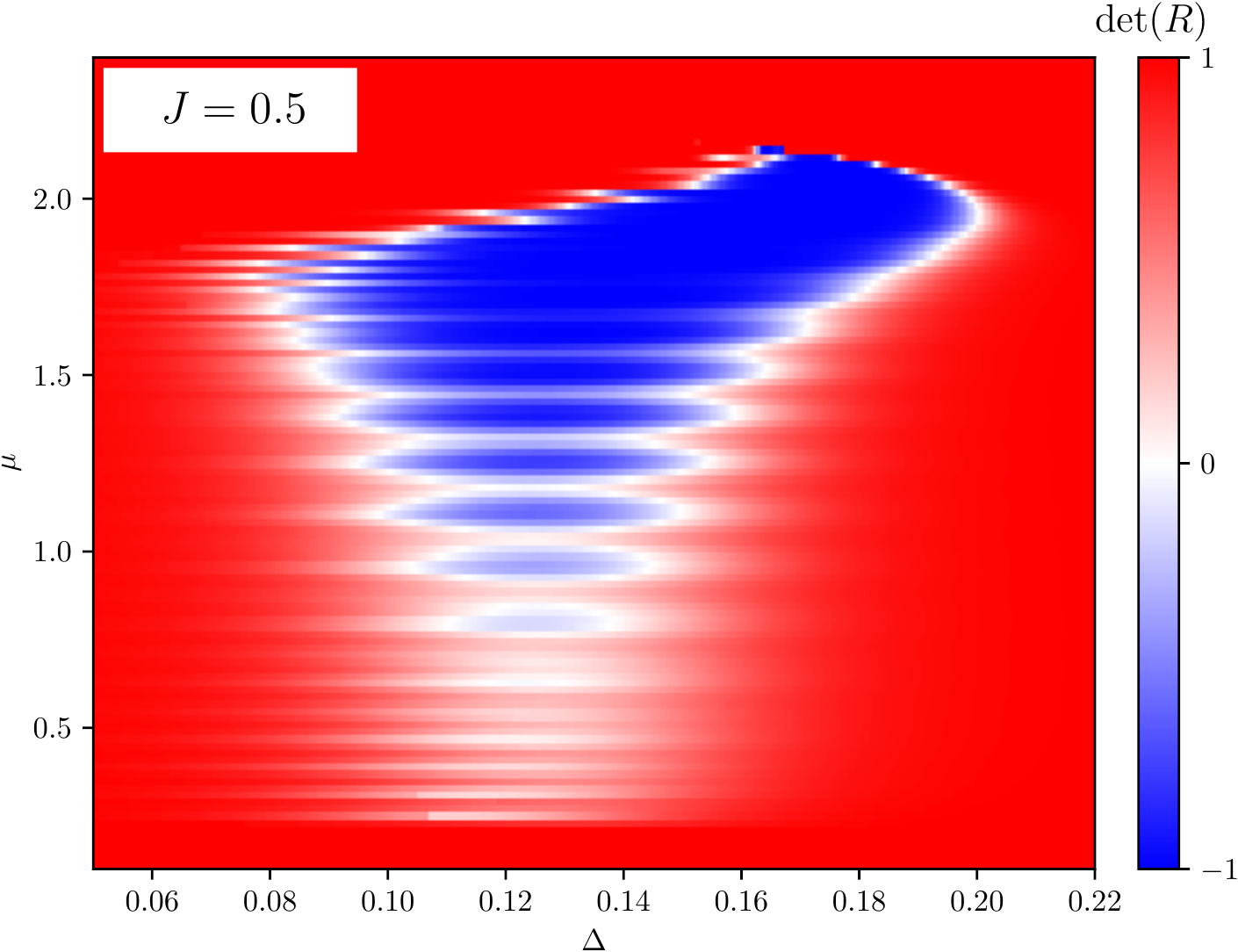}
    \includegraphics[width=0.44\textwidth]{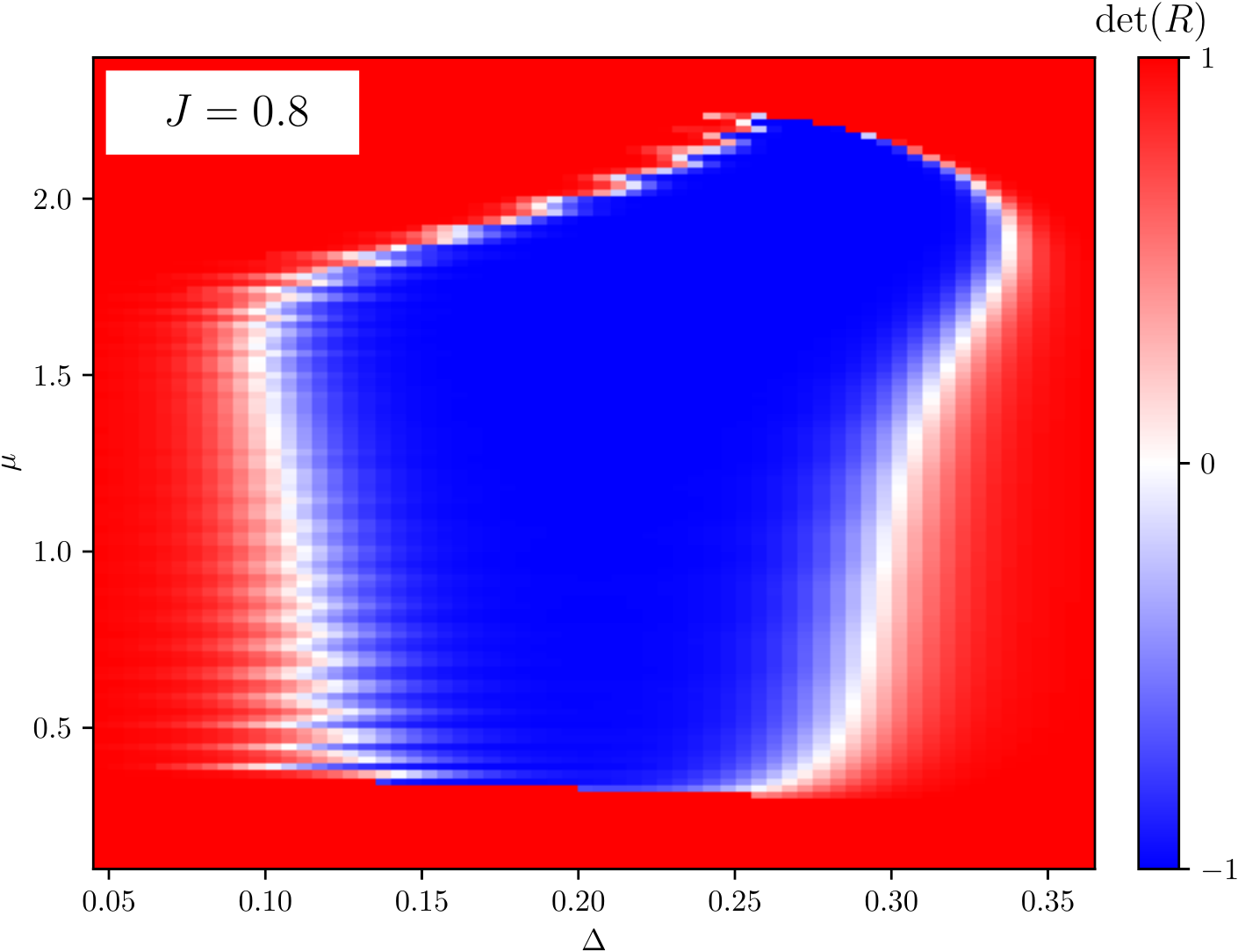}
    \includegraphics[width=0.44\textwidth]{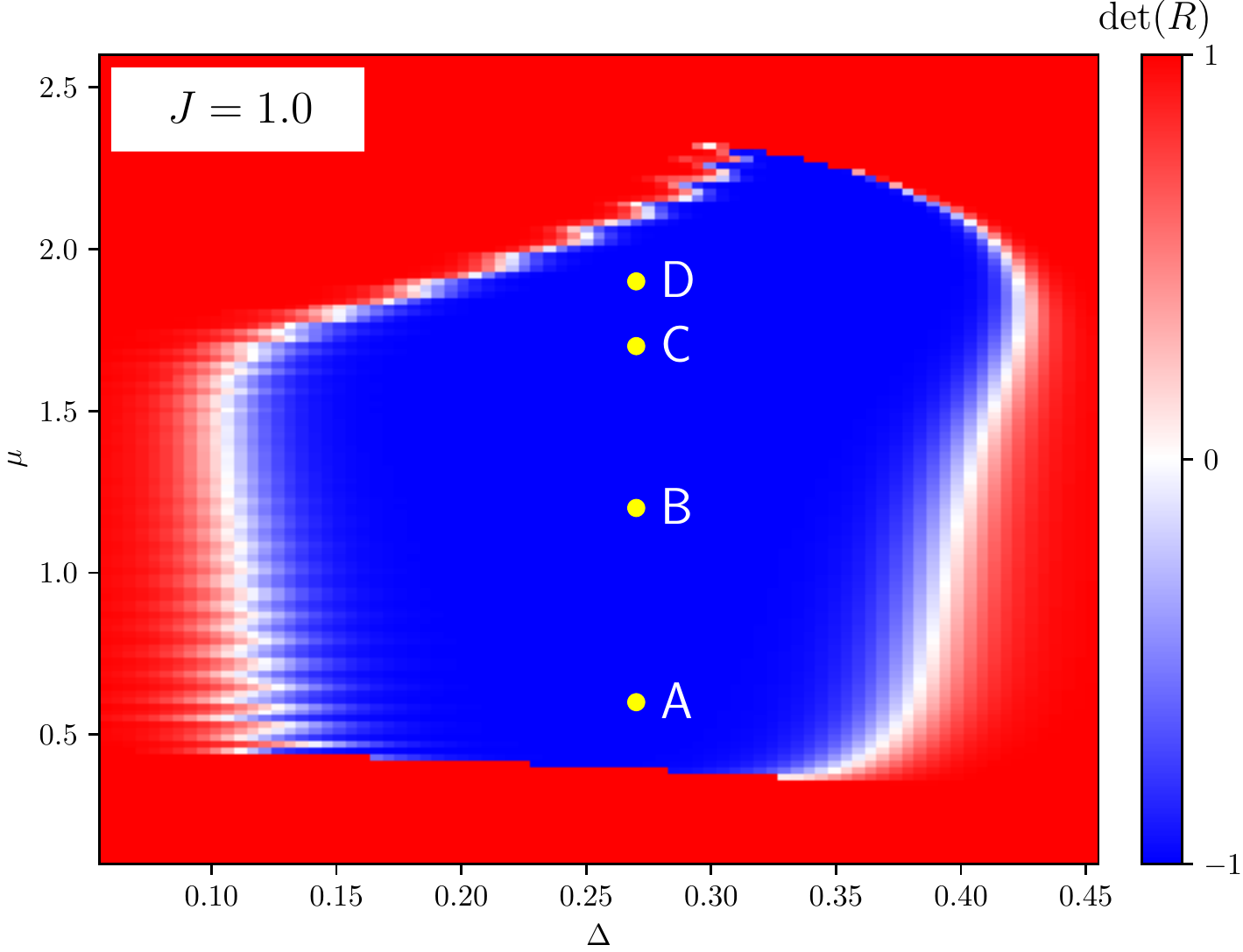}
    \includegraphics[width=0.44\textwidth]{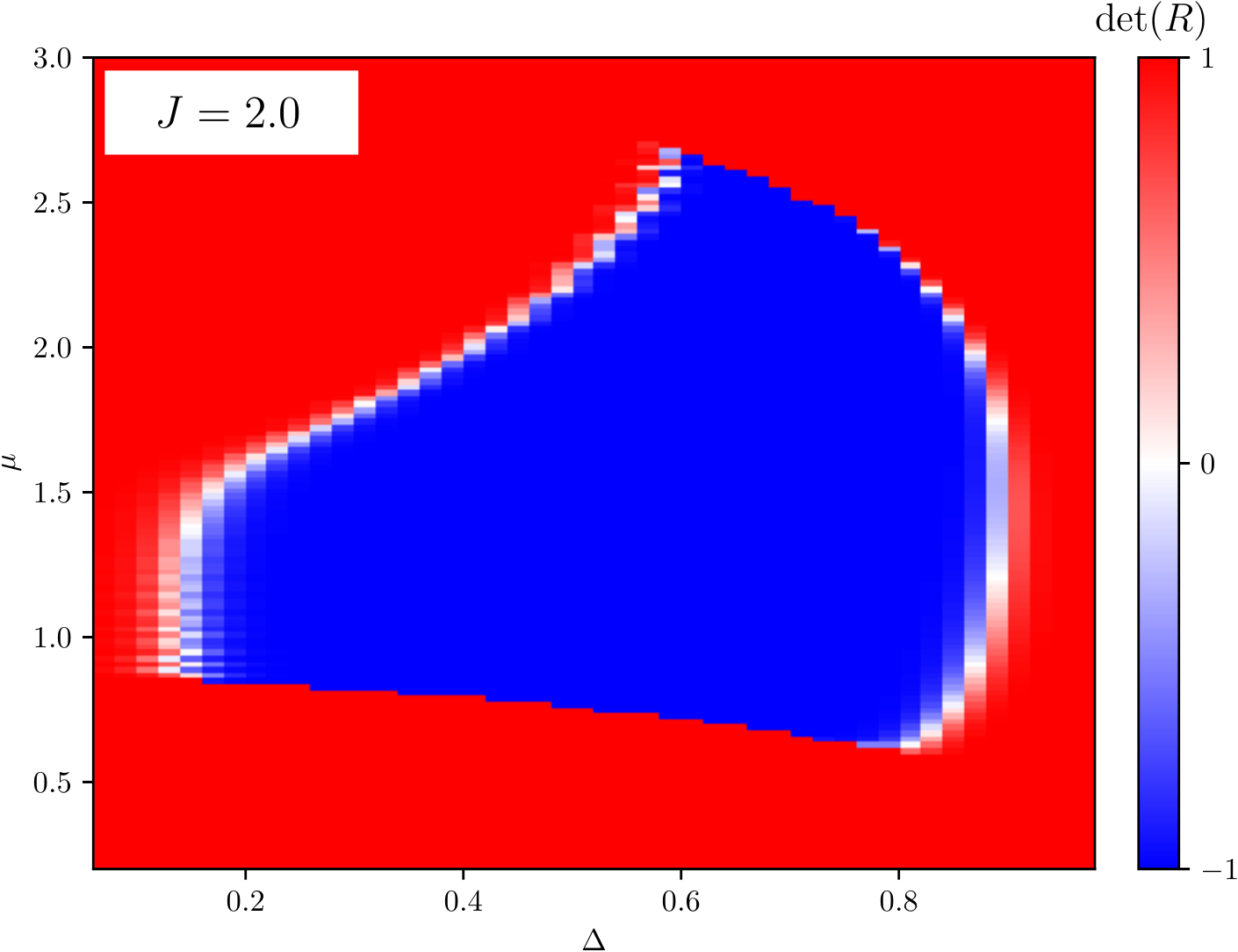}
    \includegraphics[width=0.44\textwidth]{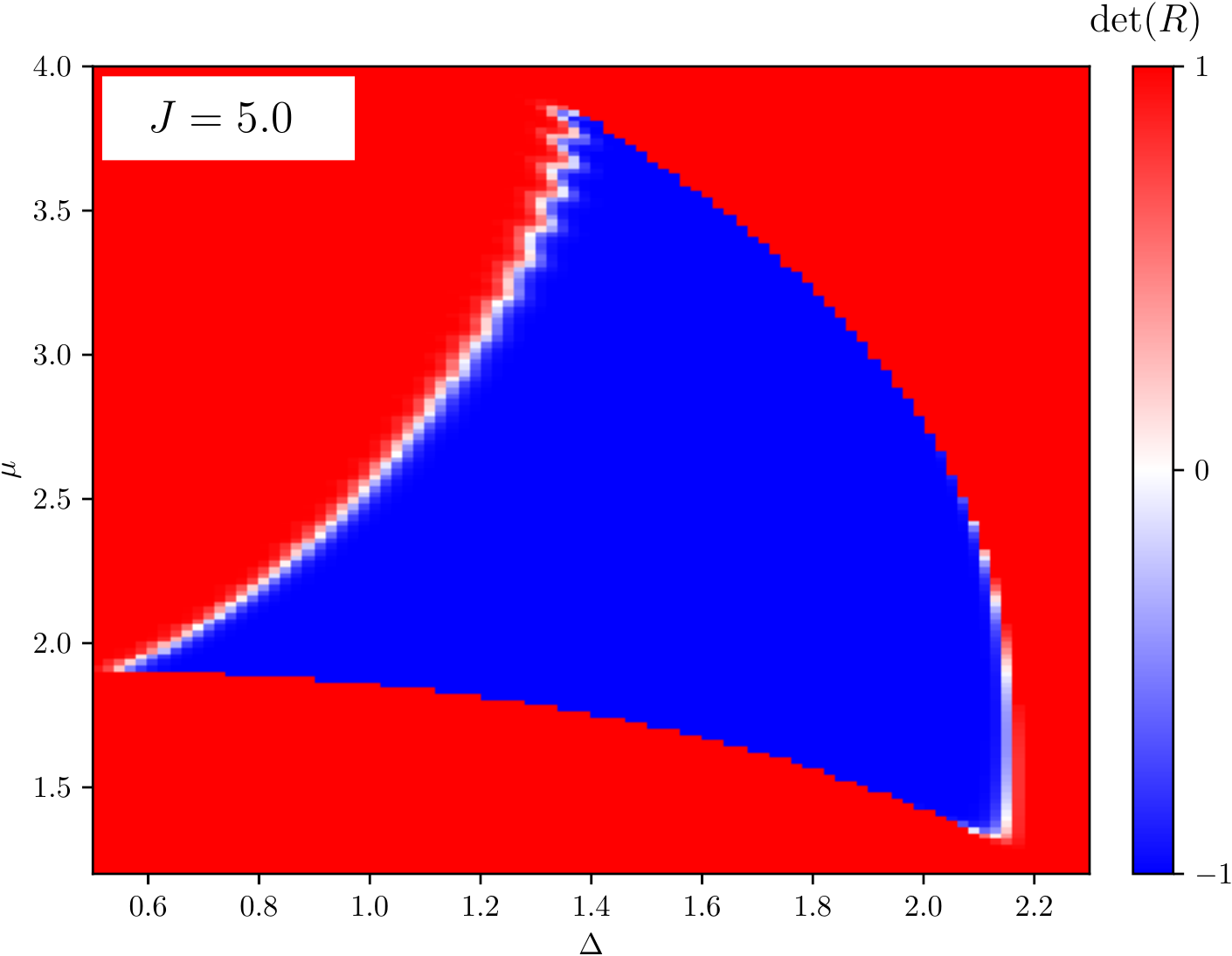}
    \includegraphics[width=0.44\textwidth]{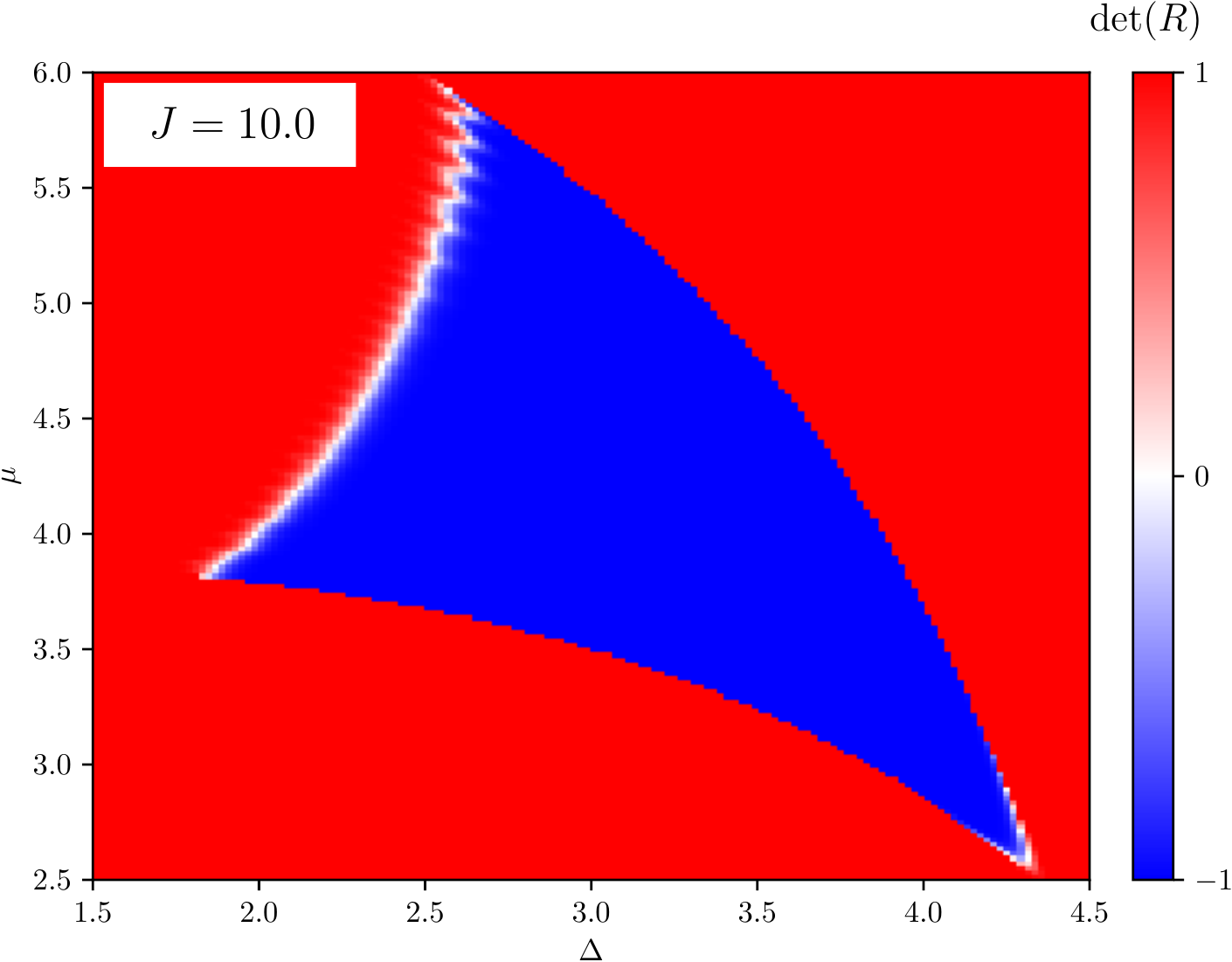}
    \caption{Zero temperature value of $\det(R)$ [see Eq. (\ref{topo_number})] as a function of $\Delta$ and $\mu$ for different values of $J$, ranging from 0.5 to 10. The blue regions represent the topologically nontrivial phase ($Q<0$) with the Majorana end-modes. We have chosen $q$ to minimize the ground state energy for the chain of $L=70$ lattice sites. Yellow circles labelled A, B, C, D indicate the parameters, for which results are presented in Fig.~\ref{fig:topo_temp}. \label{fig:topo_ph_diags}}
\end{figure*}

Ground state of the Hamiltonian (\ref{eq:hamil}) refers to some characteristic pitch $q=q_{*}$, which is determined from minimization of its energy. Since in 1D metals the static spin susceptibility diverges at $2k_F$, where $k_F$ is the Fermi momentum, it has been suggested that also in presence of the proximity--induced pairing the system will self--organize into a helical structure with the spiral pitch $q_{*}$ coinciding with the  momentum $2k_F$~\cite{Simon2013,klinovaja.stano.13,Vazifeh.Franz.13}. 
However, even in absence of the induced superconductivity the spiral pitch that minimizes 
the ground state energy can deviate from $2k_F$ if one goes beyond the Born approximation in the RKKY scheme\cite{Schecter2015}. 
We have investigated numerically variation of the ground state energy with respect to the model parameters and found, that $q_*\approx 2k_F$ only in some regimes, whereas generally $q_*$ can vary from 0 (fully polarized magnetic moments) to $\pi/a$ (anitferromagnetic ordering). Fig.~\ref{fig:gs_en}a shows a typical example of the ground state energy dependence on $q$.

To distinguish the trivial from nontrivial superconducting phases we have computed the topological number ${\mathbb Z}_2$, determining it from the scattering matrix~\cite{akhmerov.11,fulga.11}. Here we follow the procedure described in Ref.~\onlinecite{choy.11}. We have thus computed the scattering matrix $S$ of the chain
  \begin{equation}
    S=\left(\begin{array}{cc}
      R & T'\\
      T & R'
    \end{array}\right),
    \label{matrix_s}
  \end{equation}
  where $R$ and $T$ ($R'$ and $T'$) are $4\times 4$ reflection and transmission matrices at the left (right) side of the chain. This matrix (\ref{matrix_s}) describes transport through the chain
  \begin{equation}
    \left(\begin{array}{c}\psi_{-,{\rm L}} \\ \psi_{+,{\rm R}}\end{array}\right)
    =S\left(\begin{array}{c}\psi_{+,{\rm L}} \\ \psi_{-,{\rm R}}\end{array}\right),
  \end{equation}
  where $\psi_{\pm,L/R}$ are the right or left moving modes ($\pm$) at the left or right edge ($L/R$) at the Fermi level. The topological quantum number is given by~\cite{choy.11}
  \begin{equation}
      {\cal Q}={\rm sign\,det}(R) = {\rm sign\,det}(R').
      \label{topo_number}
  \end{equation}
  The scattering matrix $S$ can be obtained from multiplication of the individual transfer matrices of all the lattice sites. Since the product of numerous transfer matrices is numerically unstable, we converted them into a composition of the unitary  matrices,  involving only eigenvalues of unit absolute value.

The spiral pitch $q$ can in general be treated as an independent parameter and we can study the topological properties of the Hamiltonian (\ref{eq:hamil}) as its function. Fig.~\ref{fig:topo_qmin_J=2} shows $\det(R)$ versus $q$ and $\Delta$ for $J=2$ and several values of the chemical potential $\mu$  (analogous data have been obtained by us also for the stronger coupling $J$). In each panel we display the spiral pitch $q_{*}$ (yellow line), that minimizes the ground state energy. 
Such curves resemble the results obtained previously in the weak coupling limit $J$
(see Fig.~3 in Ref.~\onlinecite{Paaske2016a}).
Let us remark, that for the wide range of model parameters the spiral pitch $q_*(\Delta)$ indeed coincides with the topological region. It means that the system has a natural tendency towards self-adjusting the local magnetic moments in a way that guarantees the topologically nontrivial superconducting state~\cite{Vazifeh.Franz.13,Simon2013,Paaske2016}. Nevertheless, closer inspection of Fig.~\ref{fig:topo_qmin_J=2} reveals  that such tendency is not universal. For instance, for $\mu=0.5$ the topological region does not overlap with $q_*$. Also, for $\mu=2.5$ the topological state exists for $0.08 \lesssim \Delta \lesssim 0.87$, but it coincides with $q_{*}$ only in a narrow regime $0.52 \lesssim \Delta \lesssim 0.72$. 
Fig.~\ref{fig:topo_ph_diags} shows examples of the topological phase diagrams with respect to $\Delta$ and $\mu$ for the nanochain consisting of $70$ sites, assuming the stable spiral orderings $q=q_{*}$. Role of the finite--size effects is presented in Appendix \ref{app:diagrams} (Fig.~\ref{fig:fss}). We noticed that with increasing length $L$ the topological regions gradually expand and their boundaries become sharper.

\section{Role of thermal effects}
\label{sec:fin-temp}

Influence of finite temperatures on the model (\ref{eq:hamil}) can be seen in a twofold way: by thermal broadening of the Fermi--Dirac distribution function of itinerant electrons and by disturbance induced among the classical local moments ${\bm S}_i$. Since the energy resulting from rearrangement of the magnetic moments is much lower than costs of the thermal excitations of itinerant electrons, we focus on  fluctuations of the classical moments and assume that fermions are in their ground state\cite{MacDonald2001}. Such fluctuations are expected to suppress ordering of the local moments, indirectly affecting the topological superconducting phase.

To estimate the critical temperature $T_{c}$ up to which the topologically nontrivial state can persist, we have performed the MC simulations for the localized magnetic moments. Since the Hamiltonian~(\ref{eq:hamil}) includes  both the quantum (fermions) and classical (localized magnetic moments) degrees of freedom, we apply the method used in Ref.~\onlinecite{maska_czajka.06}. At each MC step a randomly chosen localized magnetic moment is rotated, the Hamiltonian (\ref{eq:hamil}) with actual configuration of ${\bm S}_i$ is diagonalized and the trial move is accepted or rejected according to the Metropolis criterion based on the free energy instead of the internal energy. During such routine we have computed the topological quantum number ${\cal Q}$ and various correlation functions. Great advantage of the MC method is that we do not need any particular ansatz for the magnetic order what is crucial for inspecting the self--organized structures composed of, e.g., several coexisting phases\cite{Scalettar2015}.

Most of our results refer to the magnetic moments confined to a plane, 
therefore only the azimuthal angles $\phi_i$ have been varied in MC simulations.  Sec.~\ref{sec:3Dspiral} presents some results for the case when this constraint is relaxed. In what follows, we discuss the most interesting results obtained within the aforementioned  algorithm.

\subsection{Correlation function}
\label{sec:correlation_length}

\begin{figure}[b!]
    \centering
\includegraphics[width=0.48\textwidth]{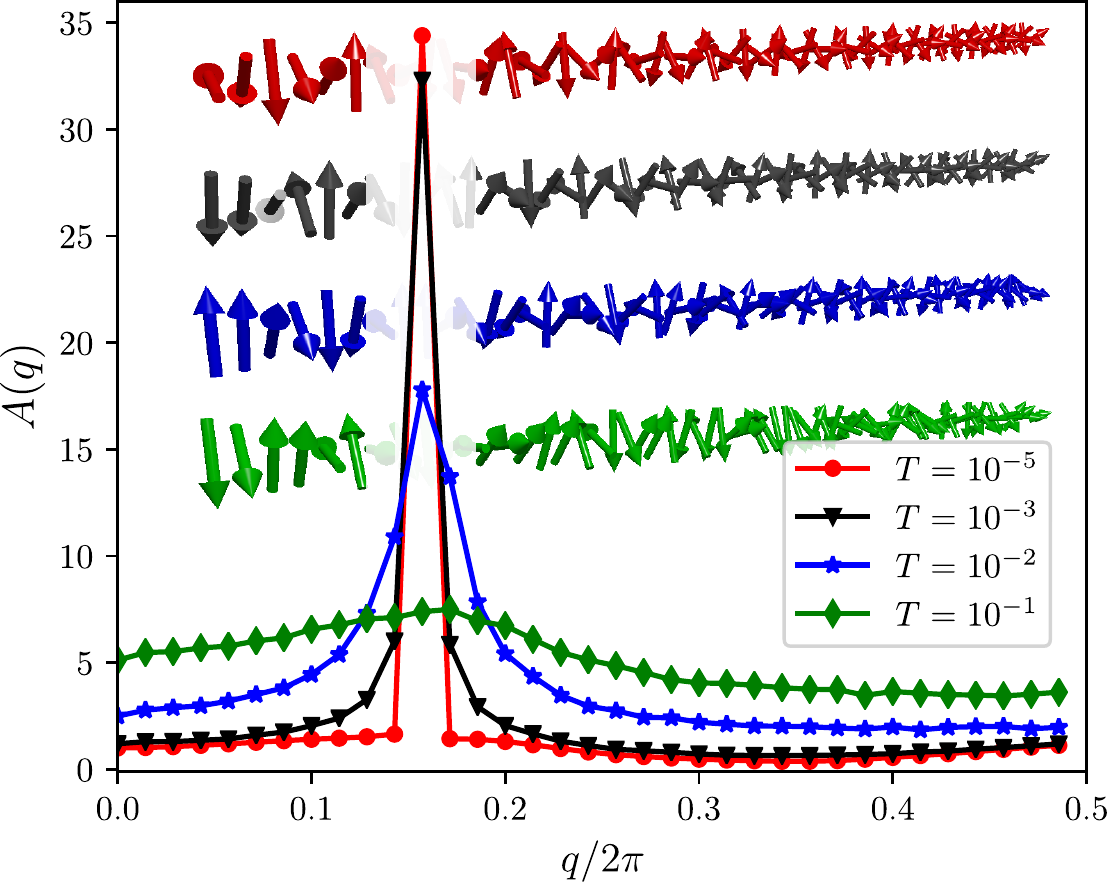}
\caption{The structure factor of the magnetic order obtained for $J=1$ and the model parameters referring to the point C in Fig.~\ref{fig:topo_ph_diags}. Results are  averaged over $10^5$ statistically independent configurations generated during MC runs at temperatures $T=10^{-5},\,10^{-3},\,10^{-2}$ and $10^{-1}$. The arrows (whose colors correspond to the Fourier transforms in the main) show representative configurations at various temperatures.
\label{fig:fft}}
\end{figure}

In Sec.~\ref{sec:TZero} we have inspected the long--range spiral ordering of the ground state. Here, we analyze how this order is affected by thermal fluctuations. In Fig.~\ref{fig:fft} we show  the structure factor of the magnetic order $A(q)= {1/L\sum_{jk}e^{iq(j-k)}\langle {\bm S}_j\cdot{\bm S}_k\rangle}$ obtained at different temperatures, as indicated. At very low temperature there is a narrow peak at $q=q_{*}$, indicating that magnetic configurations are nearly identical with the perfect zero--temperature long-range order. With increasing temperature this peak remains at its original position, but its width substantially broadens and its height is reduced. This signals that thermal fluctuations are detrimental for the magnetic ordering. We illustrate this behavior in the inset in Fig.~\ref{fig:fft}, where spatial configurations of ${\bm S}_j$  are presented for indicated temperatures.

Stability of the spiral order against thermal fluctuations is determined by the strength and range of the effective interaction between the localized magnetic moments. The interaction is mediated by itinerant electrons which are paired  through the proximity effect. Since the long--range type of the RRKY interaction in one--dimensional systems results from the gapless nature of excitations near the Fermi point, it is possible that in our case the effective interaction can differ from the standard one typical for metals. Proximity to a bulk superconductor can substantially affect its range, which should be important for any  magnetic order at finite temperatures\cite{Mermin_Wagner}. In particular, if the interaction varies as $r^{-\alpha}$ the long--range order could exist for $\alpha<2$ in the one--dimensional classical spin--$S$ Heisenberg model\cite{Rogers1981,Froehlich1978}.

\begin{figure}[b]
    \centering
\includegraphics[width=0.48\textwidth]{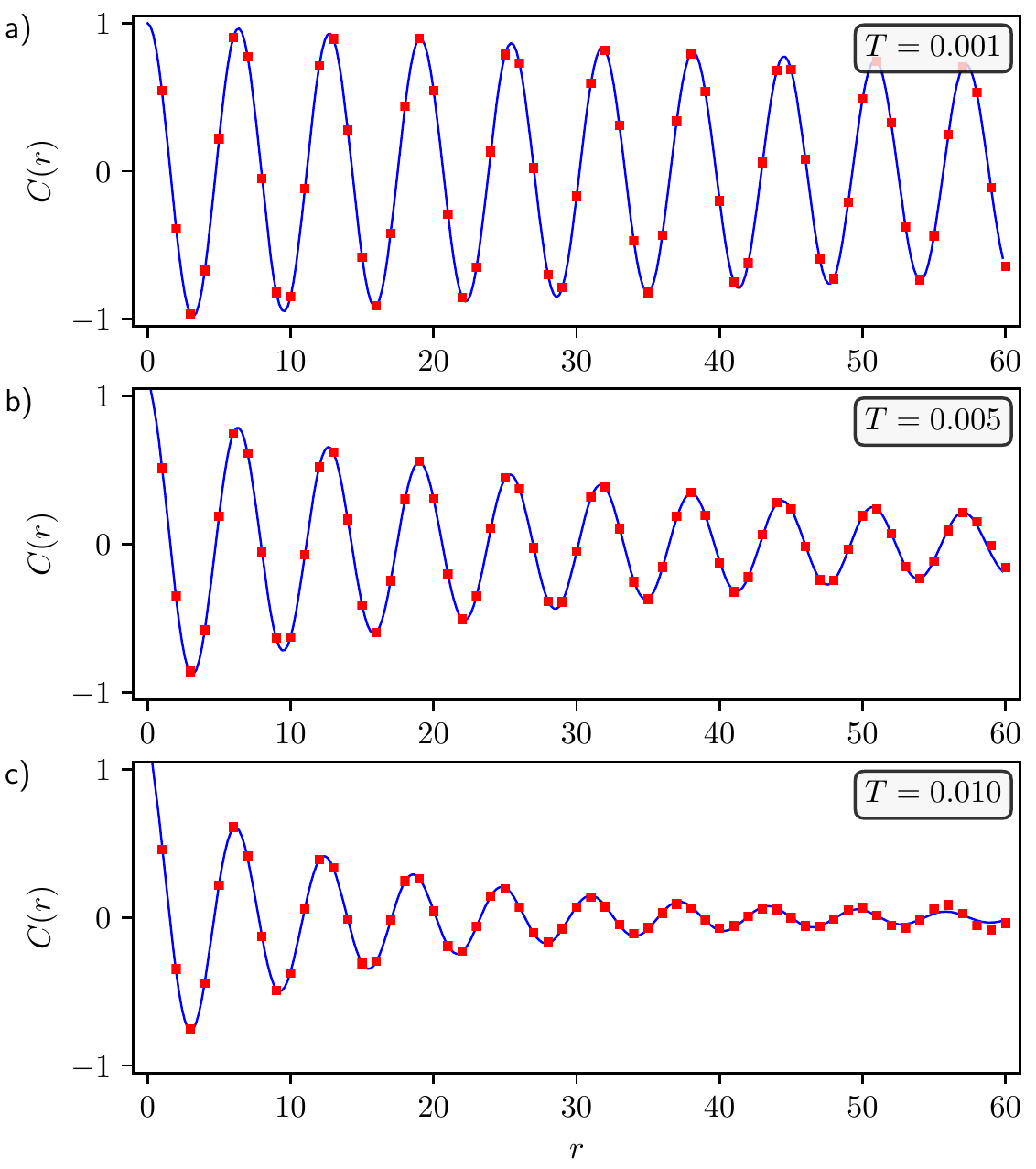}
\caption{Correlation function between the local magnetic moments (\ref{corr_funct}) as a function of distance $r$ obtained at representative temperatures for $\mu=1.7$. The red thick points show
the MC data while the blue line is the best fit with a function defined in  Eq.~(\ref{eq:MC_fit}).
\label{fig:corr_fun}}
\end{figure}

\begin{figure}
    \centering
\includegraphics[width=0.48\textwidth]{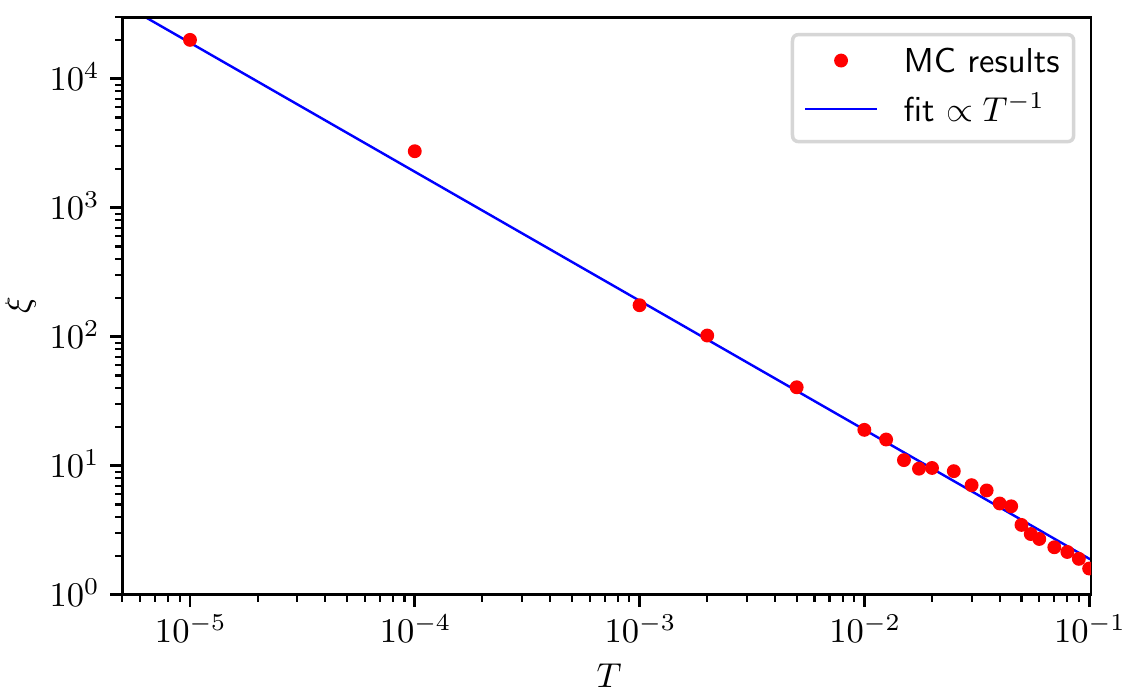}
\caption{Log--log plot of the correlation length versus temperature for the same model parameters as in Fig.~\ref{fig:corr_fun}. The red thick dots display the MC data and  the blue line is the best fit with a function $\xi(T)=A \; T^{-1}$.
\label{fig:corr_len}}
\end{figure}

To get an insight into effective interactions between the localized moments and role of the thermal effects we have analyzed the correlation function defined as
\begin{equation}
C(r)=\frac{1}{L-r-2s}\sum_{i=s}^{L-r-s} \langle {\bm S}_i\cdot{\bm S}_{i+r}\rangle ,
\label{corr_funct}
\end{equation}
where $L$ denotes the nanochain length and $s$ is a small offset introduced to minimize the finite size effects. Results of our numerical MC computations obtained for three representative temperatures are presented by the thick red dots in Fig.~\ref{fig:corr_fun}. 
The simulations show that the exponential decay of the two--point correlation function has a power law correction. The classical Ornstein--Zernike power $(d-1)/2$, where $d$ is the dimensionality of the system, vanishes in a one-dimensional system\cite{cardy_1996}. Here, however, the MC results can be very well fitted by
\begin{equation}
C(r) \propto \cos(qr)\:r^{-\alpha} \: e^{-r/\xi(T)} ,
\label{eq:MC_fit}
\end{equation}
where $\alpha$ is small ($\alpha\ll 1$) and slightly temperature dependent. Major influence of thermal fluctuations is seen by the correlation length $\xi(T)$ (see Fig.\ \ref{fig:corr_len}). We have also treated $q$ as a fitting parameter, but it turned out that 
even at elevated temperatures its value was very close to $q_*$ that minimizes the ground state energy.

Fitting the MC results by $C(r)$ defined in Eq.~(\ref{eq:MC_fit}) has enabled us to determine the temperature-dependent correlation length. As can be seen in  Fig.~\ref{fig:corr_len}, it diverges for $T\to 0$, indicating that the effective interaction is too short--ranged to produce any long--range order at finite temperatures. Nevertheless, at sufficiently low temperature the correlation length is comparable to the nanowire length therefore the system remains in the topologically nontrivial state with the Majorana modes located at its edges. For unambiguous verification of such possibility we have directly calculated the topological properties of the system at finite temperatures (Sec.~\ref{sec:finite_topo}).

\subsection{Topological phase at finite temperatures}
\label{sec:finite_topo}

\begin{figure}[htb]
    \centering
    \includegraphics[width=0.5\textwidth]{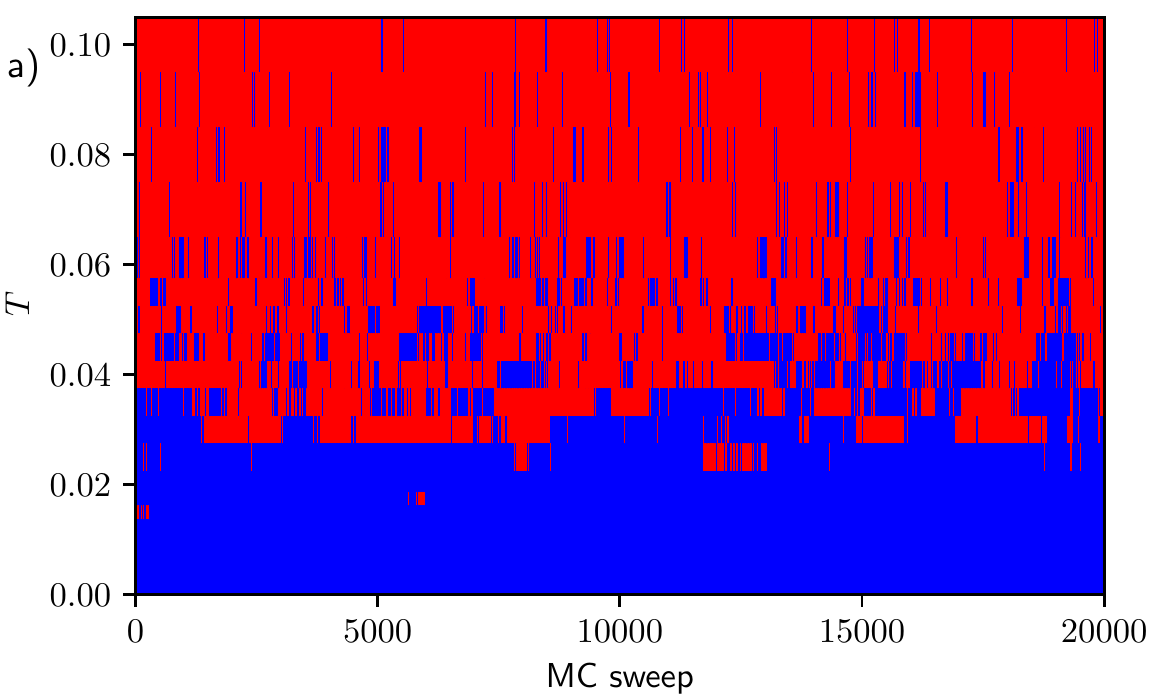}
    \includegraphics[width=0.48\textwidth]{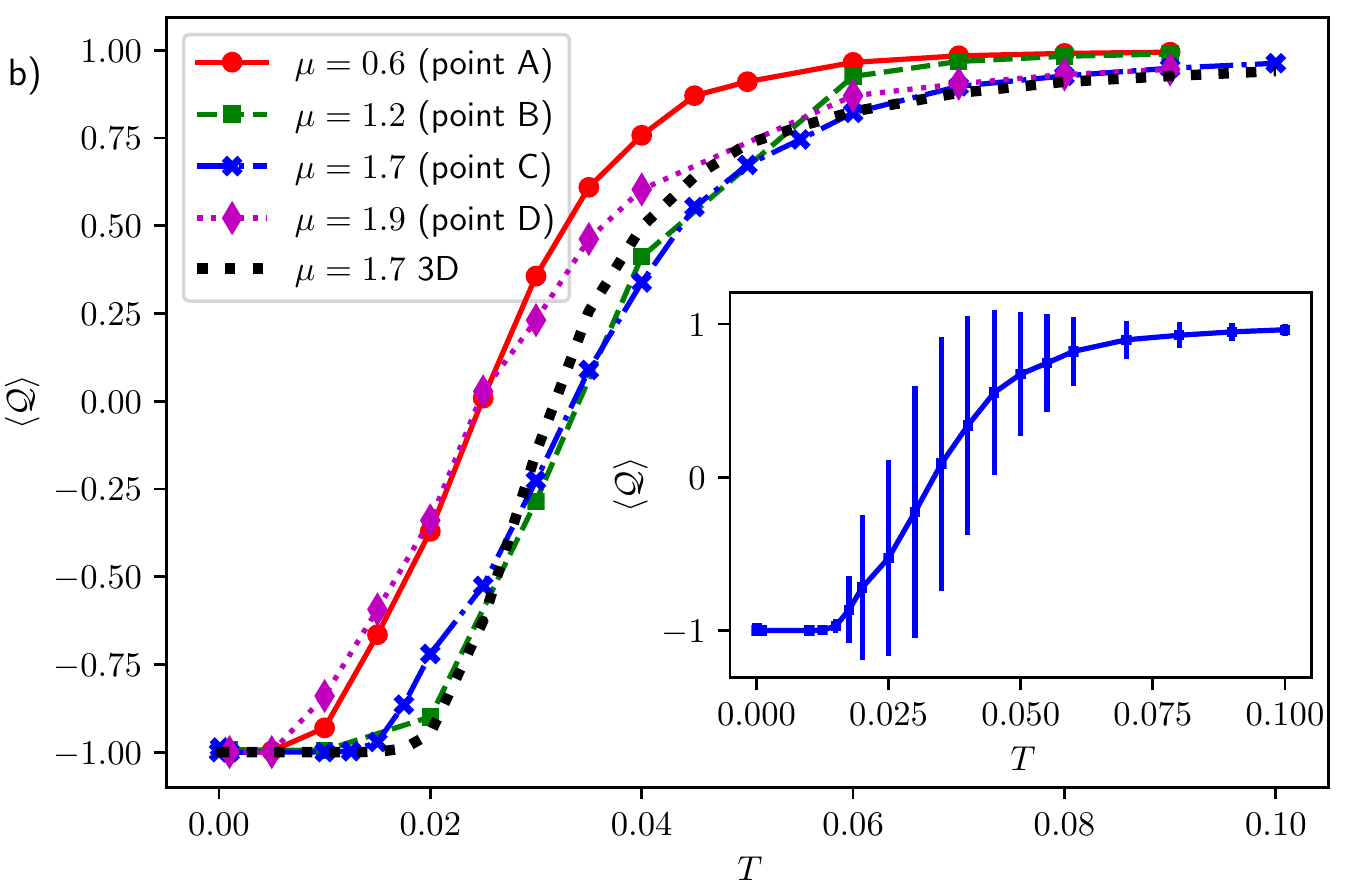}
    \caption{a) Variation of topological invariant ${\cal Q}$ during the MC sweeps obtained for varying temperature. Blue regions correspond to ${\cal Q}=-1$ and red to ${\cal Q}=+1$, respectively. The results refer to $J=1,\:\mu=1.7,\:\Delta=0.27$  (point C in Fig.~\ref{fig:topo_ph_diags}). b) Temperature dependence of the invariant ${\cal Q}$ averaged over $10^5$ MC sweeps for the model parameters indicated by points A-D in Fig.~\ref{fig:topo_ph_diags}.  The thick black dotted line marked as ``3D'' shows ${\cal Q}$ calculated for point C under the assumption that the magnetic moments ${\bm S}_i$ are not confined to a plane (see Section \ref{sec:3Dspiral}). The inset presents the standard deviation of ${\cal Q}$ obtained for point C.} 
    \label{fig:topo_temp}
\end{figure}

Fig.~\ref{fig:topo_temp}a displays variation of the topological invariant ${\cal Q}$ during the MC runs performed at different temperatures (vertical axis). We have chosen the model parameters which guarantee the system to be in topologically nontrivial phase at zero temperature (point C in Fig.~\ref{fig:topo_ph_diags}, corresponding to $J=1.0$). At this point the system is in its topologically nontrivial state for chains of different lengths (see Fig.~\ref{fig:fss}). 
We clearly notice that with increasing temperature more and more frequently the system prefers the topologically trivial state. 
Such gradual changeover from the topological to non-topological phase depends on the chemical potential (Fig.~\ref{fig:topo_temp}b) and other parameters as well. Roughly speaking, for the chosen set of model parameters the topological phase exists up to the critical temperature  $T_{c} \sim 0.05$ (in units of the hopping integral). Considering typical values  $t\sim 10\:$meV  \cite{Vazifeh.Franz.13,klinovaja.stano.13} this would yield the critical temperature for the topological superconducting phase $T_{c} \sim 6\:$K which is a more stringent limitation than all previous estimations~\cite{Vazifeh.Franz.13,klinovaja.stano.13,Simon2015}.

\subsection{Spectral functions}
\begin{figure*}
\centering
\includegraphics[width=0.90\textwidth]{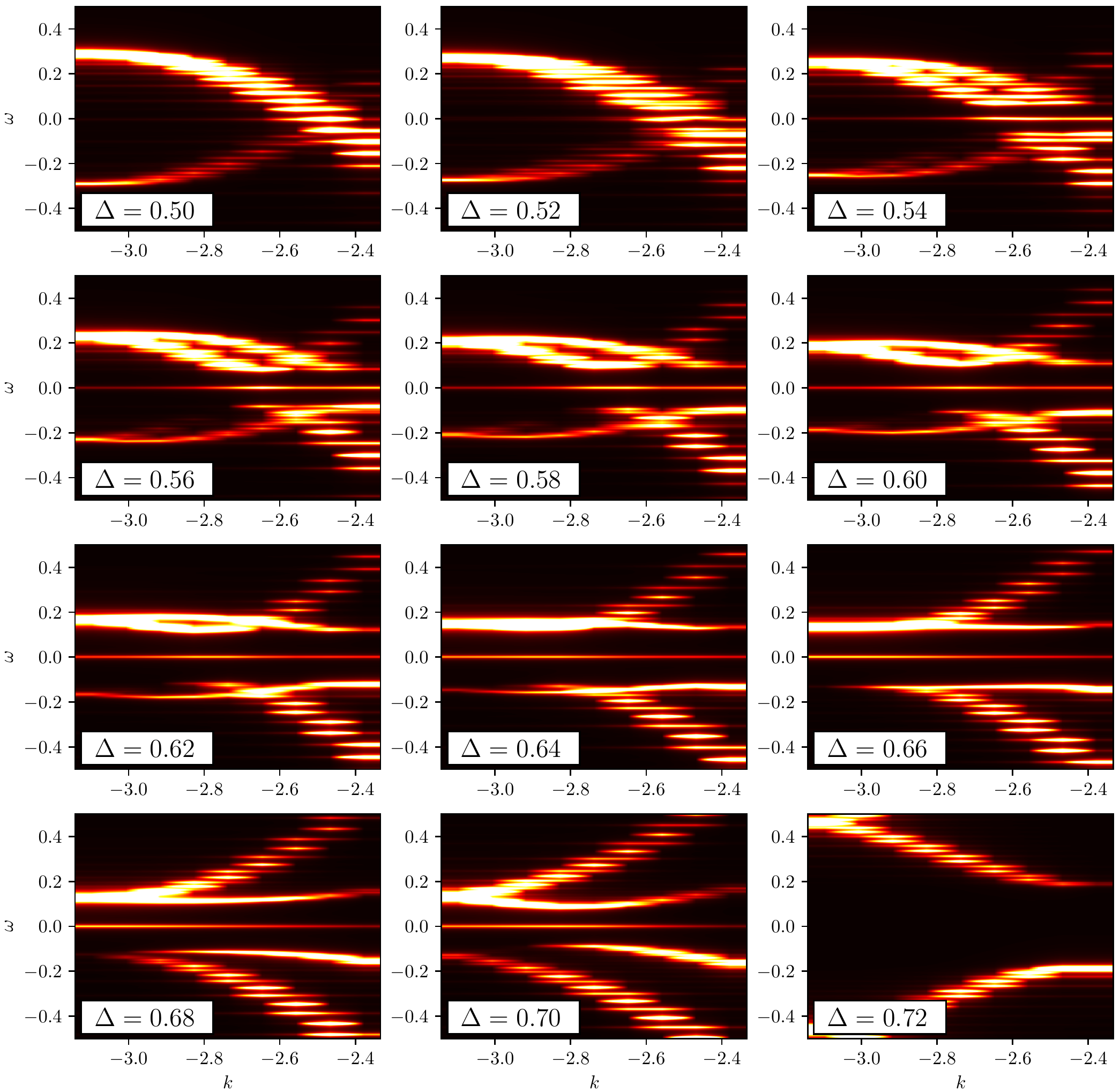}
\caption{Evolution of the zero temperature spectral functions with respect to varying $\Delta$ obtained for $q=q_{*}$ which  for the model parameter $\mu=2.5$, $J=2$ is shown by the yellow line in Fig.~\ref{fig:topo_qmin_J=2}. Note, that the presence of the zero-energy feature coincides with $q_{*}$ being in the topological region (blue area in Fig.~\ref{fig:topo_qmin_J=2}).}
\label{fig:sp_multi}
\end{figure*}

\begin{figure*}
\centering
\includegraphics[width=0.45\textwidth]{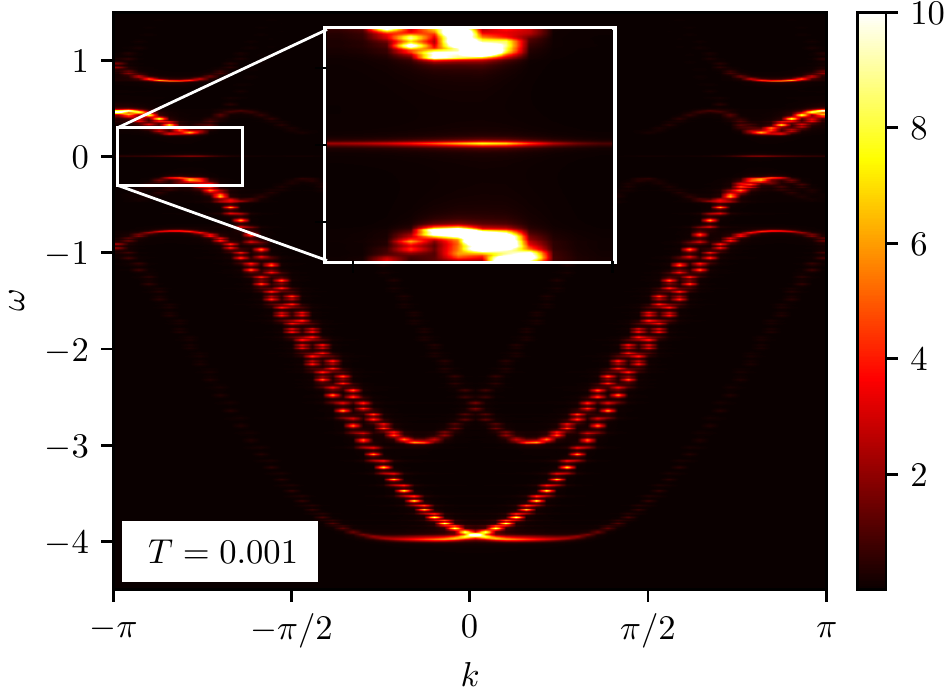}
\includegraphics[width=0.45\textwidth]{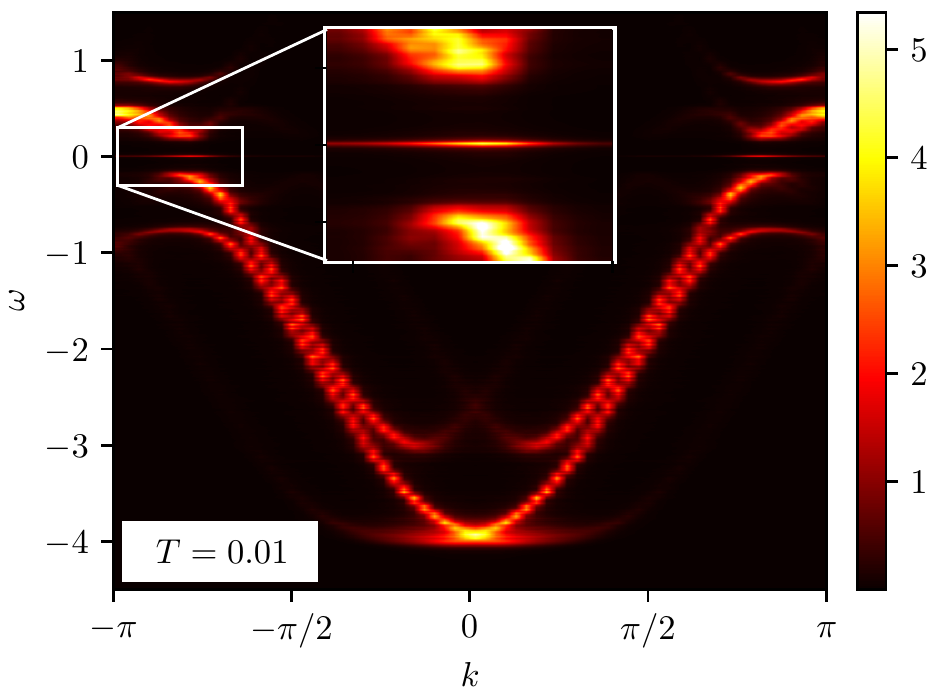}
\includegraphics[width=0.45\textwidth]{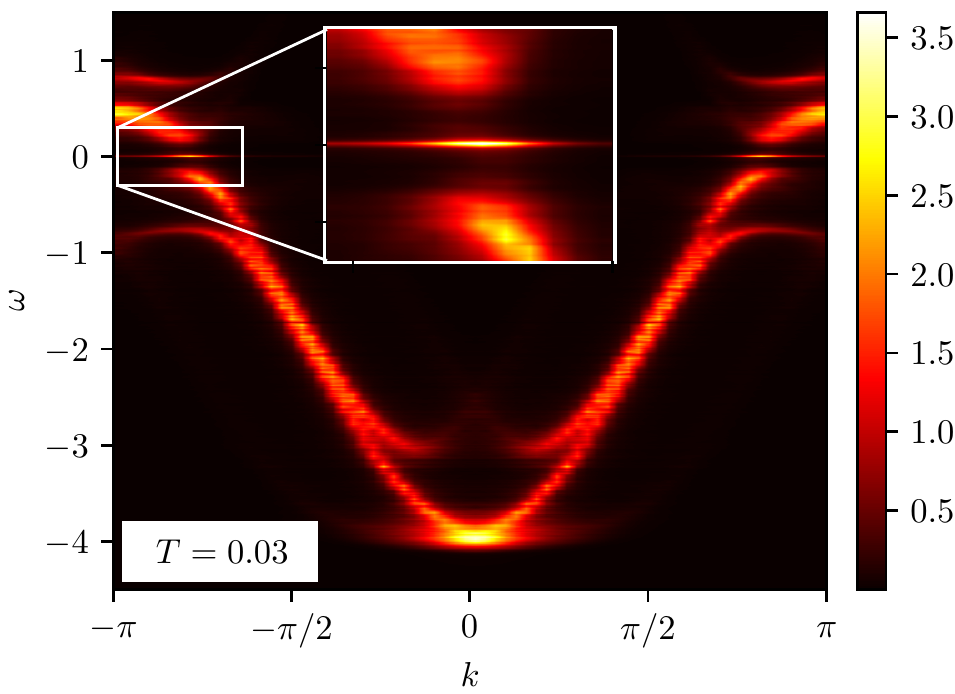}
\includegraphics[width=0.45\textwidth]{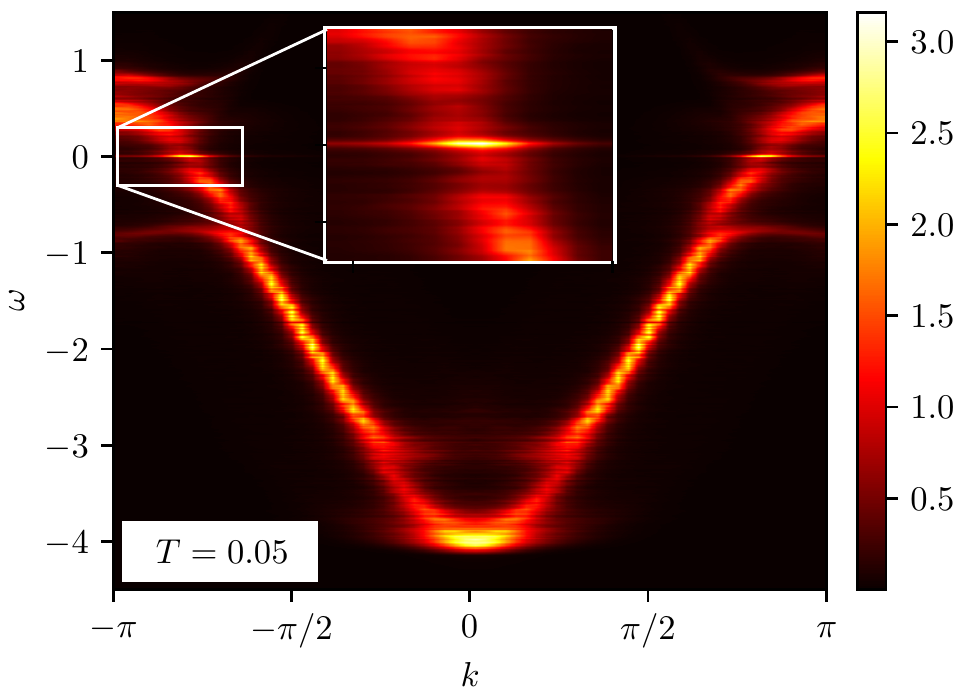}
\includegraphics[width=0.45\textwidth]{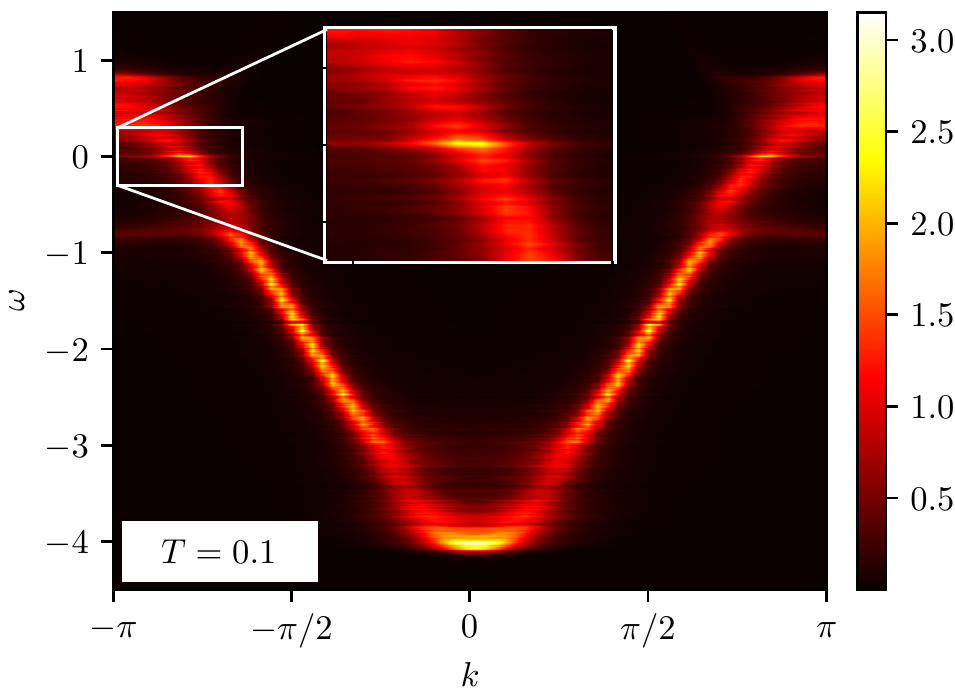}
\includegraphics[width=0.45\textwidth]{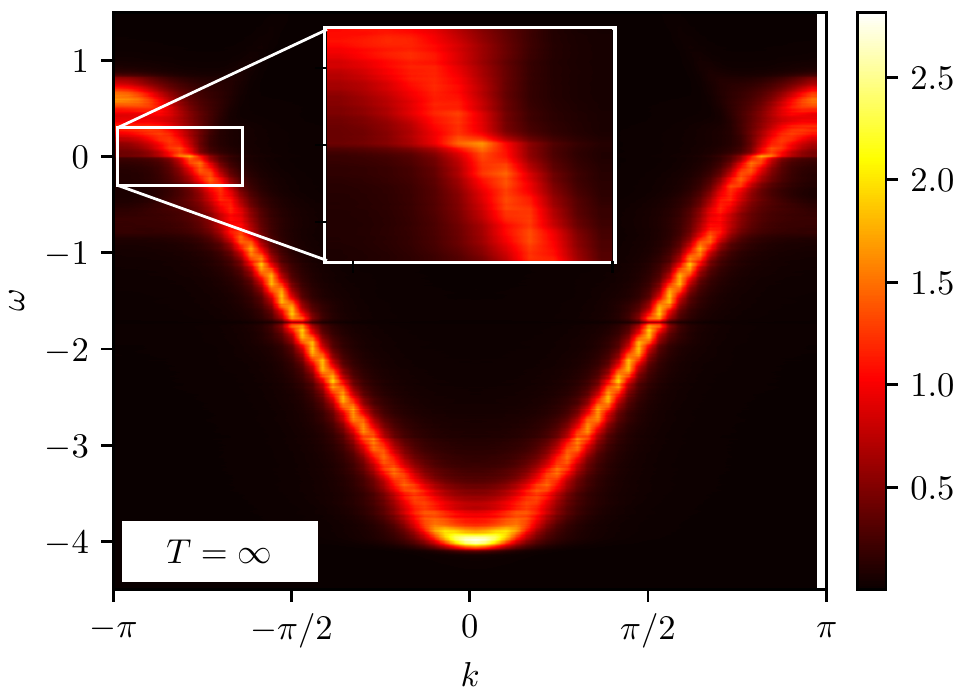}
\caption{The spectral function [defined in Eq.~(\ref{spectral_function})] averaged over $10^3$ statistically independent configurations of the localized moments $\{{\bm S}_i\}$. MC results are obtained for $J=1,\:\mu=1.7,\:\Delta=0.27$ and several temperatures, as indicated. The zoomed region displays the zero--energy mode.}
\label{fig:spectr_fun}
\end{figure*}

Another evidence for the detrimental influence of thermal effects on the topological superconductivity and the Majorana modes can be seen directly from the quasiparticle spectra of fermions. The spectral function
\begin{equation}
A(k,\omega)=-\frac{1}{\pi}{\rm Im}\:G(k,\omega+i0^+) 
\label{spectral_function}
\end{equation}
can be obtained using the single particle Green's function
\begin{equation}
G(k,z)\delta(k-k')=\sum_{m,n}\langle {\cal G}_{mn}(z)\rangle e^{i(mk-nk')} .
\label{eq:green}
\end{equation}
Here ${\cal G}_{mn}(z)=\{\left[z-H\right]^{-1}\}_{mn}$ is defined in the  real-space for  a  given  configuration of the localized moments (recall that the lattice constant $a\equiv 1$) and $\langle\ldots\rangle$ denotes averaging over configurations generated in MC runs. 

Let us first inspect the spectral function  (\ref{spectral_function})  at zero temperature to demonstrate its characteristic features upon entering the topological regime. Fig.~\ref{fig:sp_multi} presents evolution of the low energy spectrum, showing emergence of the zero-energy Majorana mode. For a given value of $\Delta$ we have computed the optimal pitch $q_{*}$ of the ground state and then determined $A(k,\omega)$ for the model Hamiltonian (\ref{eq:hamil}) with such particular configuration of the local moments $S_{i}$.  In other words, at zero temperature the averaging over configurations $\langle\ldots\rangle$ defined in Eq.~(\ref{eq:green}) was not necessary. For the chosen value $\mu=2.5$ the pitch vector $q_{*}(\Delta)$ is shown by the yellow line in Fig.~\ref{fig:topo_qmin_J=2}. In particular, we can notice the qualitative change (from topological to nontopological phase) when $\Delta$ varies from $0.70$ to $0.72$, which corresponds to the abrupt jump of $q_{*}(\Delta)$ displayed in Fig.~\ref{fig:topo_qmin_J=2}.

Influence of thermal effects of the spectral function (\ref{spectral_function}) is illustrated for the representative set of model parameters in Fig.~\ref{fig:spectr_fun}. At zero temperature the Majorana mode (appearing near boundaries of the Brillouin zone, as shown by the inset) is protected from the finite-energy Andreev quasiparticles by the topological gap. Upon increasing the temperature such topological gap gradually diminishes. This is accompanied by an ongoing disordering of the local magnetic moments leading to a broadening of all the spectral lines. Ultimately, at temperatures $T \simeq 0.05$ the topological gap is hardly visible, and the zero-energy feature merges with a continuum. Nonetheless, even at higher temperatures we could still resolve some remnants of the overdamped zero-energy mode. This brings us to the conclusion that the topological superconductivity vanishes near such critical temperature in a continuous manner (like a crossover rather than typical phase transition).

\section{Beyond coplanar ordering}
\label{sec:3Dspiral}

Finally, we have checked whether deviation of the azimuthal angle of the local moments (\ref{eq:parametrization}) from its coplanar value $\theta_i=\pi/2$ could affect the topological superconducting phase. For this purpose we have performed MC simulations, treating both angles ($\theta_i, \phi_i$) on equal footing. To find the lowest energy configuration of the localized magnetic moments we used the simulated annealing method\cite{Kirkpatrick671}. 

\begin{figure}
    \centering
    \includegraphics[width=0.38\textwidth]{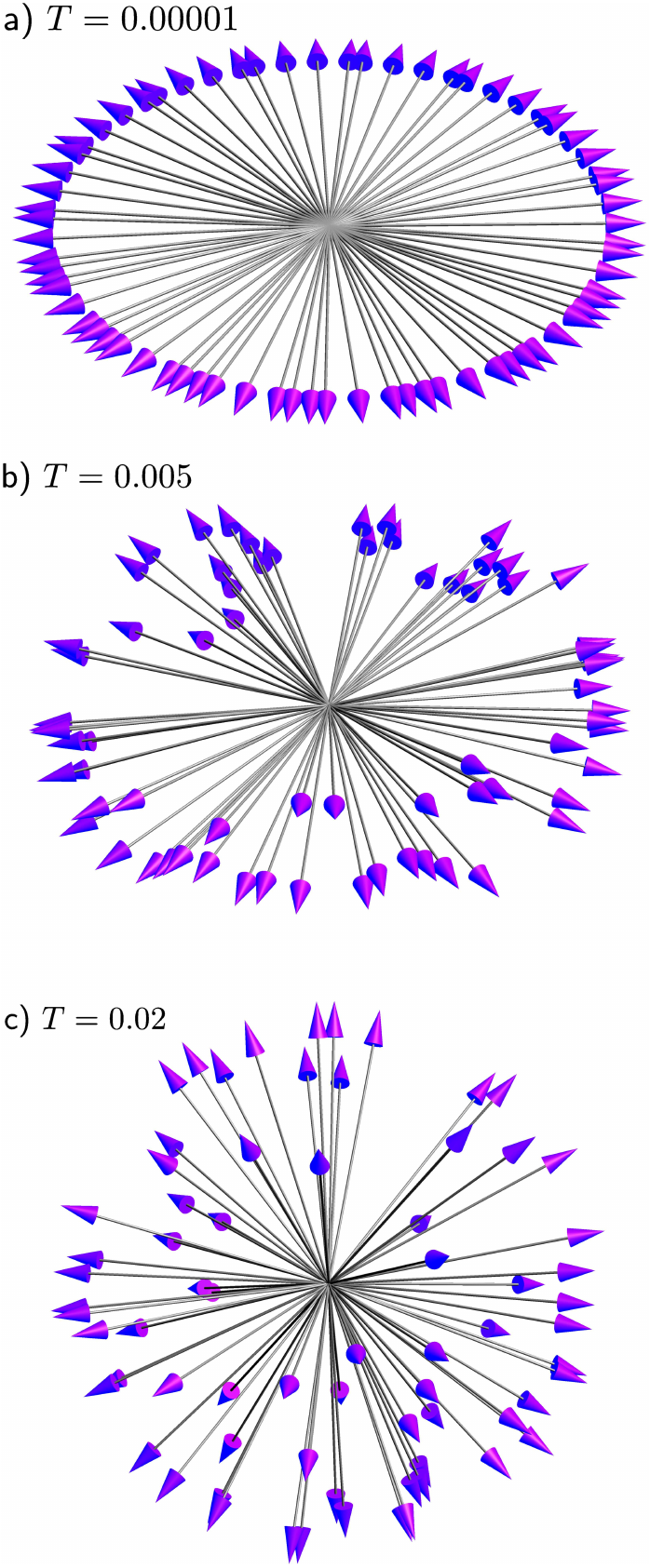}    
\caption{Orientations of the localized magnetic moments at different temperatures. The model parameters correspond
to point C in Fig.~\ref{fig:topo_ph_diags}. For the sake of clearness the origin of all these vectors has been collected to a common point and the average plane of the order tilted to be horizontal.}
\label{fig:hedgehog}
\end{figure}

At very low temperature the local magnetic moments are arranged in a coplanar spiral albeit now the plane of moments rotation is arbitrarily oriented, what reflects the symmetry of the Hamiltonian (\ref{eq:hamil}). This situation is illustrated in Fig.~\ref{fig:hedgehog}a, where the moments have been shifted so that their origins are in the same point. 
As a result, the zero temperature phase diagrams are the same as in Fig.~\ref{fig:topo_ph_diags}. With increasing temperature, the moments deviate from their coplanar arrangement (besides introducing in--plane disorder) what is illustrated in Fig.~\ref{fig:hedgehog}b and c. Similarly to the previously studied case, where the moments were confined to a plane, it may lead to destruction of the topological state. An example of such a behavior is illustrated by the thick dotted black line in Fig.~\ref{fig:topo_temp}b.
%
%
%
One can notice there that the temperature dependence of $\langle {\cal Q}\rangle$ is almost unaffected by the presence of the additional degree of freedom, what may suggest that polar angle $\theta_i$ is rather irrelevant for stability of the topologically nontrivial superconducting phase. 
 
\begin{figure}[!ht]
    \centering
    \includegraphics[width=0.45\textwidth]{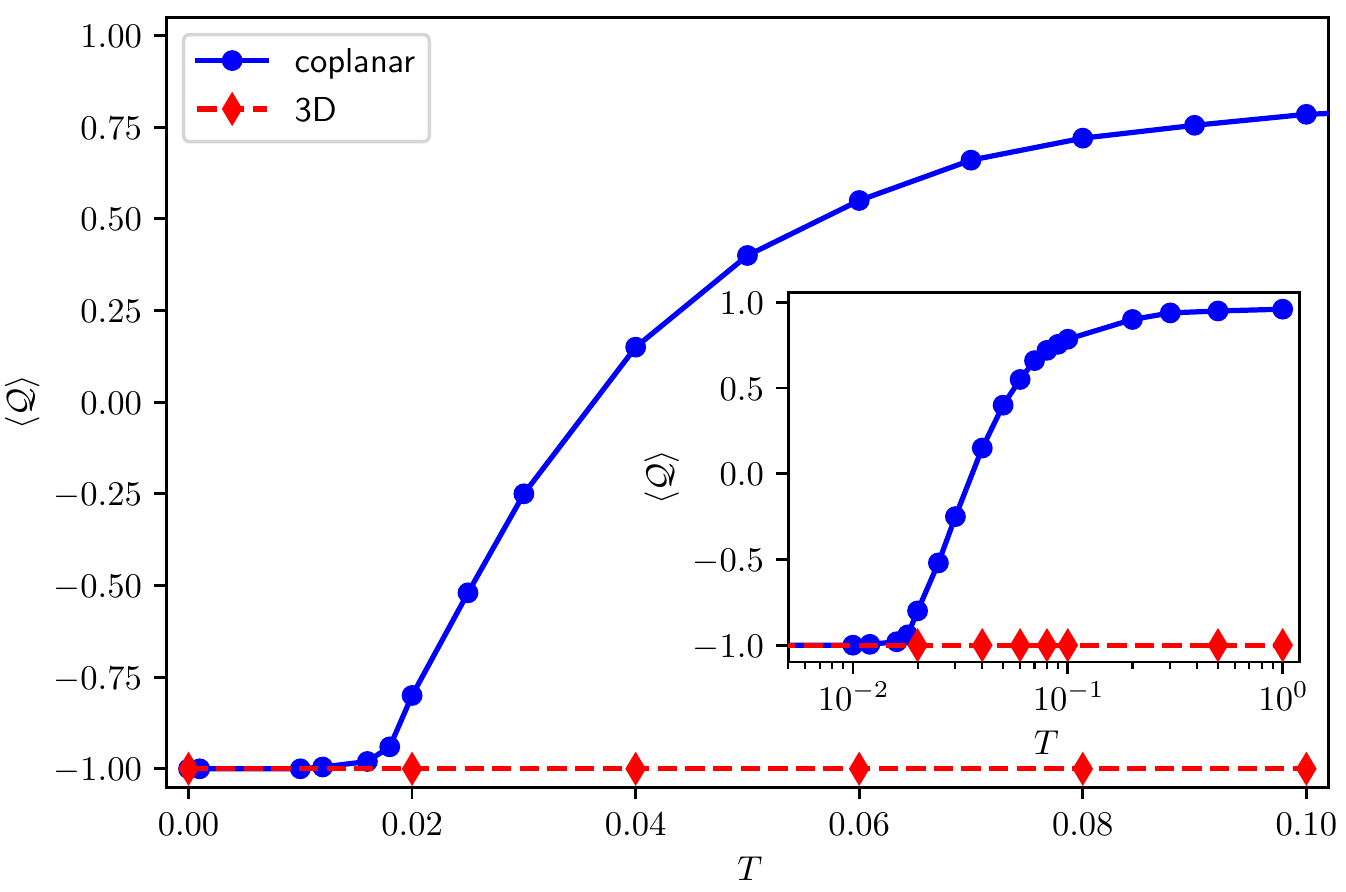}
\caption{Temperature dependence of the average topological invariant ${\cal Q}$ for coplanar configurations of the localized magnetic moments (blue solid line) and when their rotation is allowed in arbitrary direction (red dashed line marked ``3D''). The inset shows the same but in a semilogarithmic scale for a wider range of temperatures. The model parameters are $J=4,\:\Delta=0.9$ and $\mu=2.3$.}
\label{fig:topo_3D_2D}
\end{figure}

This property, however, is not universal. Fig.~\ref{fig:topo_3D_2D} shows the temperature dependence of $\langle{\cal Q}\rangle$ for a different set of the model parameters. In this case the topological phase is destroyed by increasing temperature when only in--plane thermal fluctuations of the localized moments are allowed, but it survives to pretty high temperatures when they rotate freely in all three dimensions. 
Since at high temperature the helical order vanishes, the model Hamiltonian (\ref{eq:hamil}) cannot be related to the scenario with the spin--orbit and Zeeman interactions \cite{Braunecker2010,klinovaja.stano.13}. 

However, it was shown in Ref.~\onlinecite{choy.11} that even without the helical order this Hamiltonian can have topologically nontrivial state provided the localized magnetic moments point in different directions. 
It is usually assumed that for sufficiently large $J$ the electron spin is parallel to the localized magnetic moment. We have verified this assumption by calculating the correlation function $\frac{1}{L}\sum_i \langle {\bm S}_i\cdot\hat{\bm s}_i \rangle$. The results show that the electron spin is almost completely polarized along the localized magnetic moments for {\em arbitrary} value of $J$. In such a case the Hamiltonian (\ref{eq:hamil}) can be projected onto the lowest spin band and take a form of Kitaev's chain with additional hopping to the next nearest neighbors\cite{choy.11}. In the effective Kitaev Hamiltonian the pairing potential increases with increasing 
disorder of the magnetic moments and can drive the system into the topological phase. This can explain that while the spiral ordering is destroyed at high temperature, another mechanism can still keep the system in the nontrivial state, as marked by the red line in Fig.~\ref{fig:topo_3D_2D}.

\section{Summary}
\label{sec:concl}

We have investigated stability of the topologically nontrivial superconducting phase of itinerant electrons coupled to the local magnetic moments in the finite-length nanowire proximitized to $s$-wave superconductor. We have performed the MC simulations, considering various configurations of such local moments constrained on a plane and oriented arbitrarily in all three directions. 
%
We have focused on the role played by thermal fluctuations. MC simulations clearly indicate that self-organization of the local moments into the spiral order gradually ceases upon increasing the temperature. We have found the universal scaling 
of the correlation function for the localized magnetic moments (\ref{corr_funct}) and determined the coherence length, revealing its characteristic temperature dependence  $\xi(T) \propto 1/T$. 

Our MC data for the topological invariant and analysis of the quasiparticle spectrum both unambiguously show the upper (critical) temperature  $T_{c}$, above which the topological nature of the superconducting phase no longer exists. When approaching this critical temperature from below there occurs a gradual reduction of the topological gap, protecting the zero-energy mode from the finite-energy (Andreev-type) quasiparticles, so that at $T \rightarrow T_{c}$ the Majorana modes get overdamped. Our quantitative estimations show that $T_{c} \sim 0.05$ (in units of the hopping integral) what in realistic systems would yield $T_{c} \sim 6$K. Such upper limit for existence of the topological superconducting phase could be important for  experimental and theoretical studies of the Majorana quasiparticles in the condensed matter and the ultracold atom systems. This evaluation should also be taken into account when considering future applications of the Majorana quasiparticles for quantum computing. 

The approach we used in this work is quite general and thus the model can be easily extended by taking into account other mechanisms which affect stability of the topological phase, like the spin--orbit coupling, direct interaction between the localized moments, different kinds of disorder or external magnetic field.

\acknowledgments
We acknowledge discussions with Jelena Klinovaja and Jens Paaske.
This work is supported by the National Science Centre (Poland) under the contracts DEC-2018/29/B/ST3/01892 (A.G.--G. and M.M.M.) and DEC-2017/27/B/ST3/01911 (T.D.).

\begin{figure*}
\centering
\includegraphics[width=0.44\textwidth]{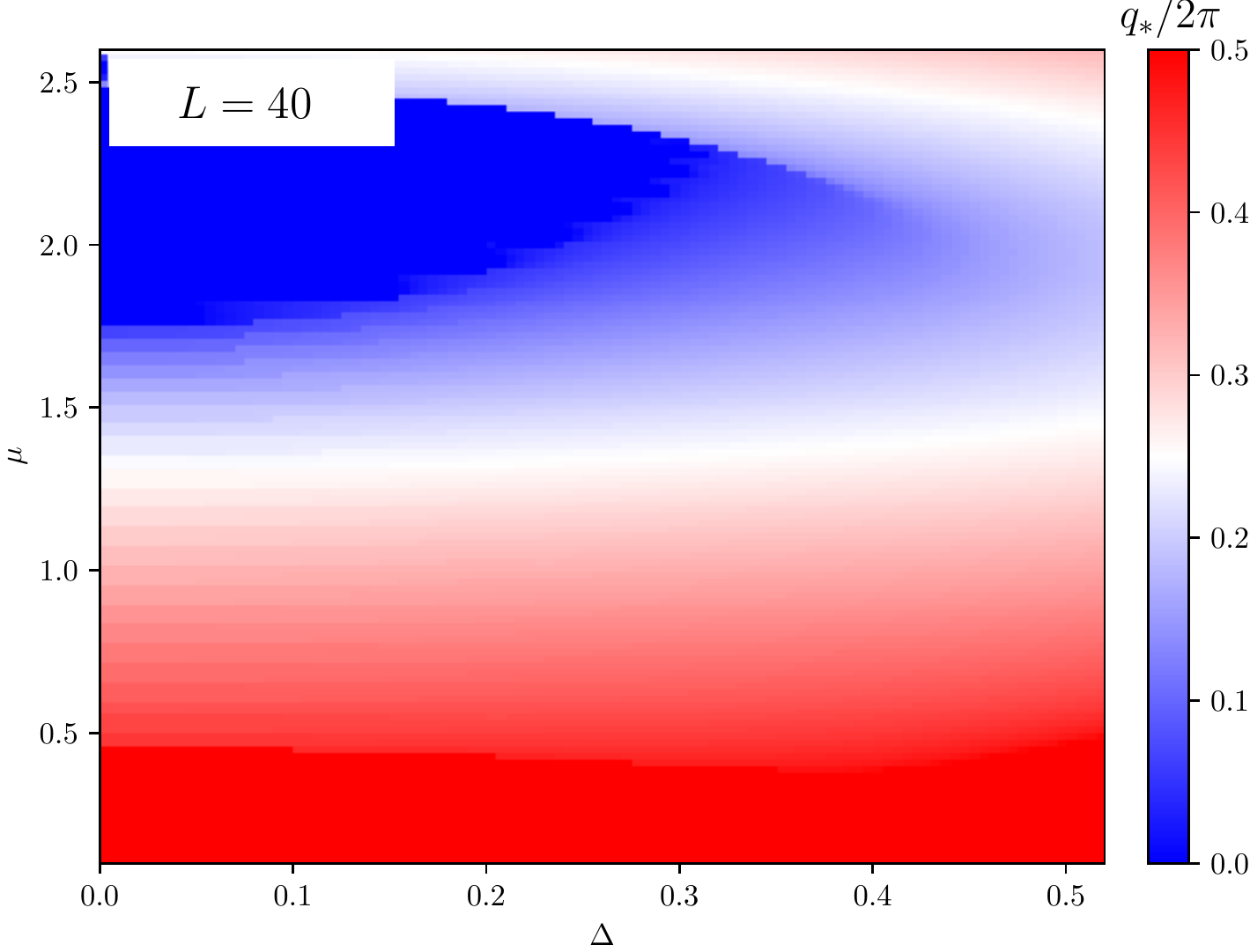}
\includegraphics[width=0.44\textwidth]{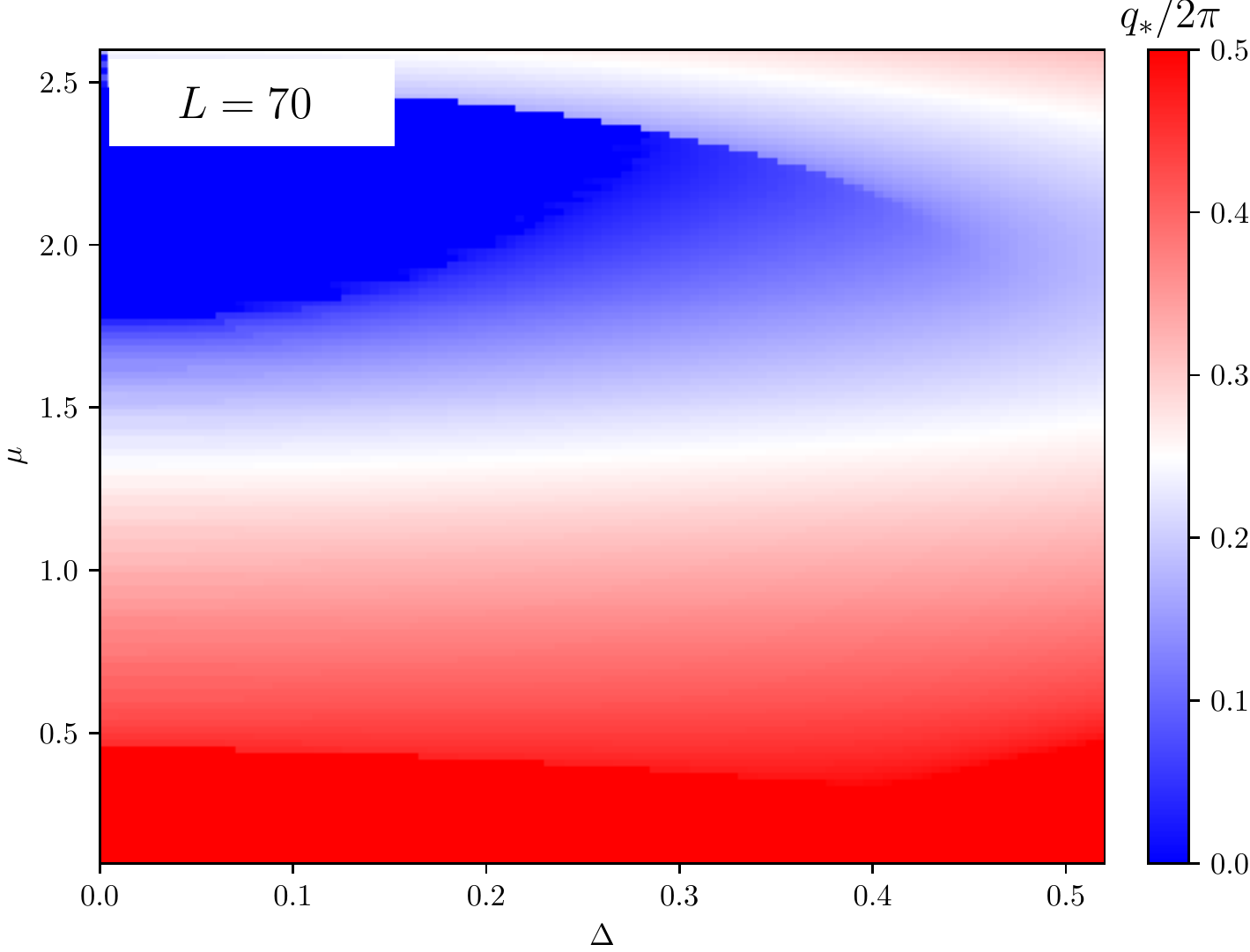}
\includegraphics[width=0.44\textwidth]{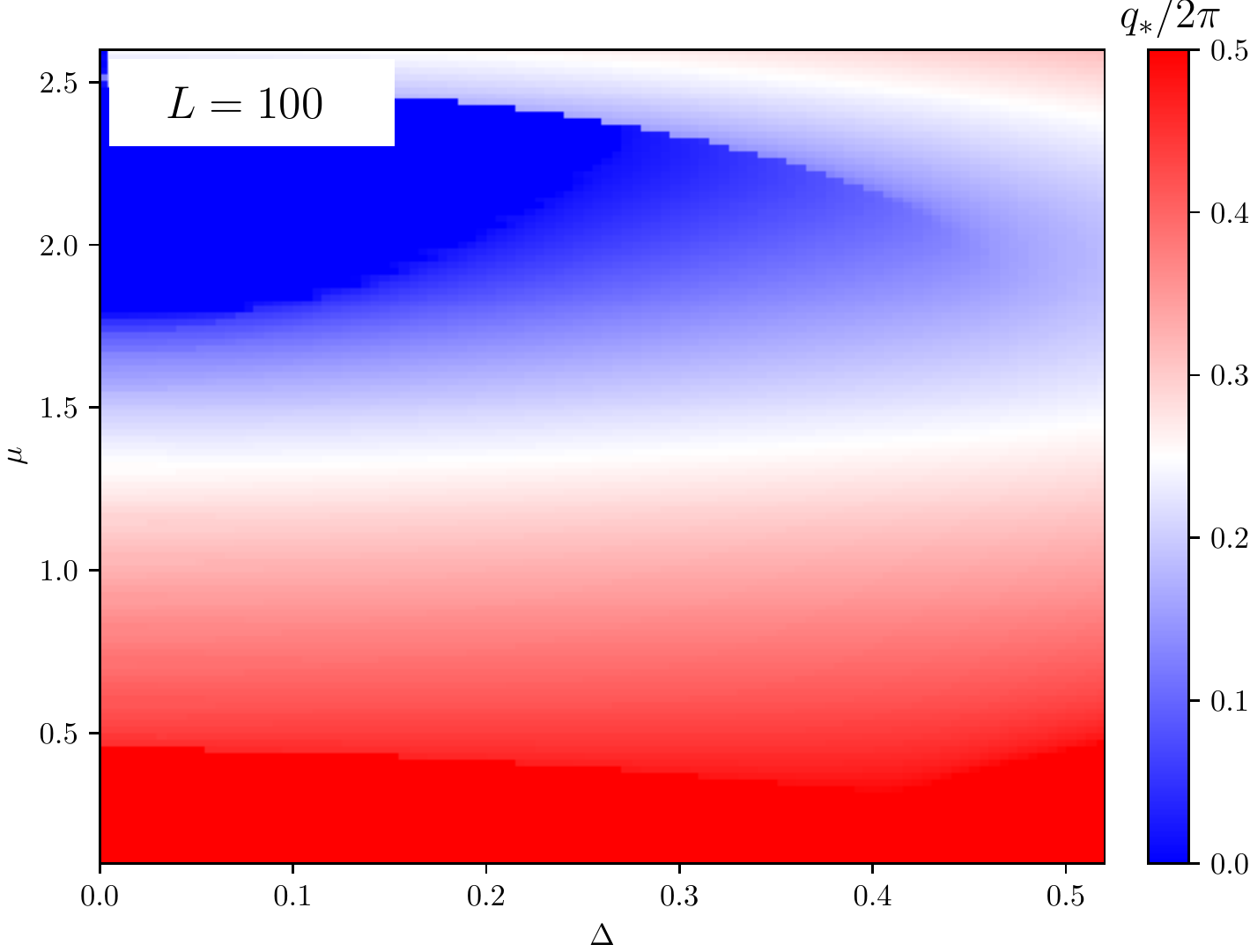}
\includegraphics[width=0.44\textwidth]{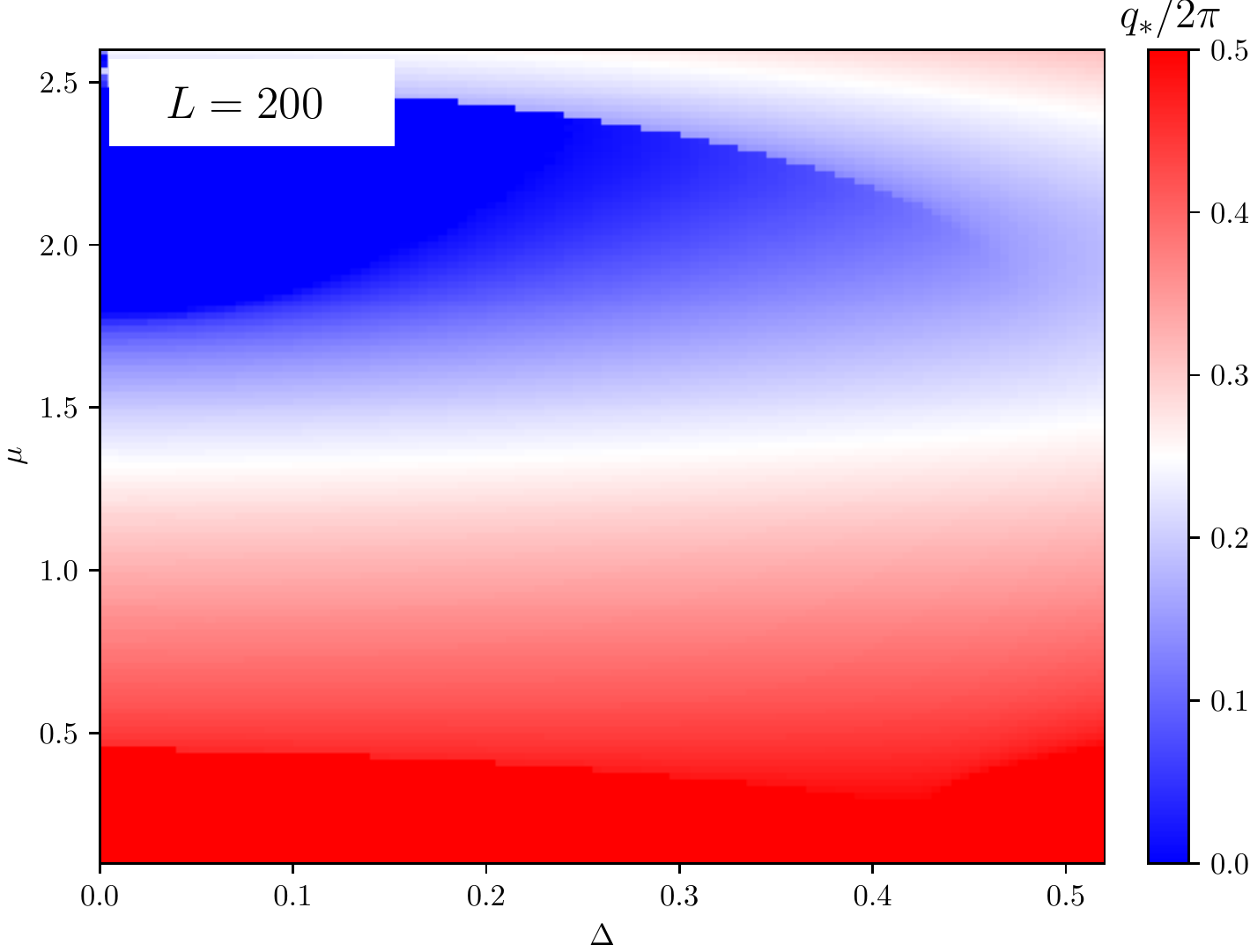}
\caption{The pitch vector $q_*$ obtained at zero temperature for $J=1$ and several sizes, 
ranging from $L=40$ to $L=200$. The model parameters are the same as in Fig.~\ref{fig:topo_ph_diags}.
\label{fig:fss_q}}
\end{figure*}

\appendix*

\section{Finite-size scaling}
\label{app:diagrams}

In the scenario based on the Rashba nanowire proximitized to a bulk superconductor a sharp transition from the topologically trivial to nontrivial regime has been predicted only for infinitely long wires and it has been emphasized~\cite{Mishmash16} that finite--size effects would smooth it into a crossover. Due to correspondence between systems with the spin--orbit and Zeeman interactions and systems with the spiral ordering of localized moments, the same effect can be expected for the present model described by the Hamiltonian ($\ref{eq:hamil}$). To verify it we performed additional calculations for various lengths $L$ of nanowires, comprising 40 to 200 sites. 

\begin{figure*}
\centering
\includegraphics[width=0.44\textwidth]{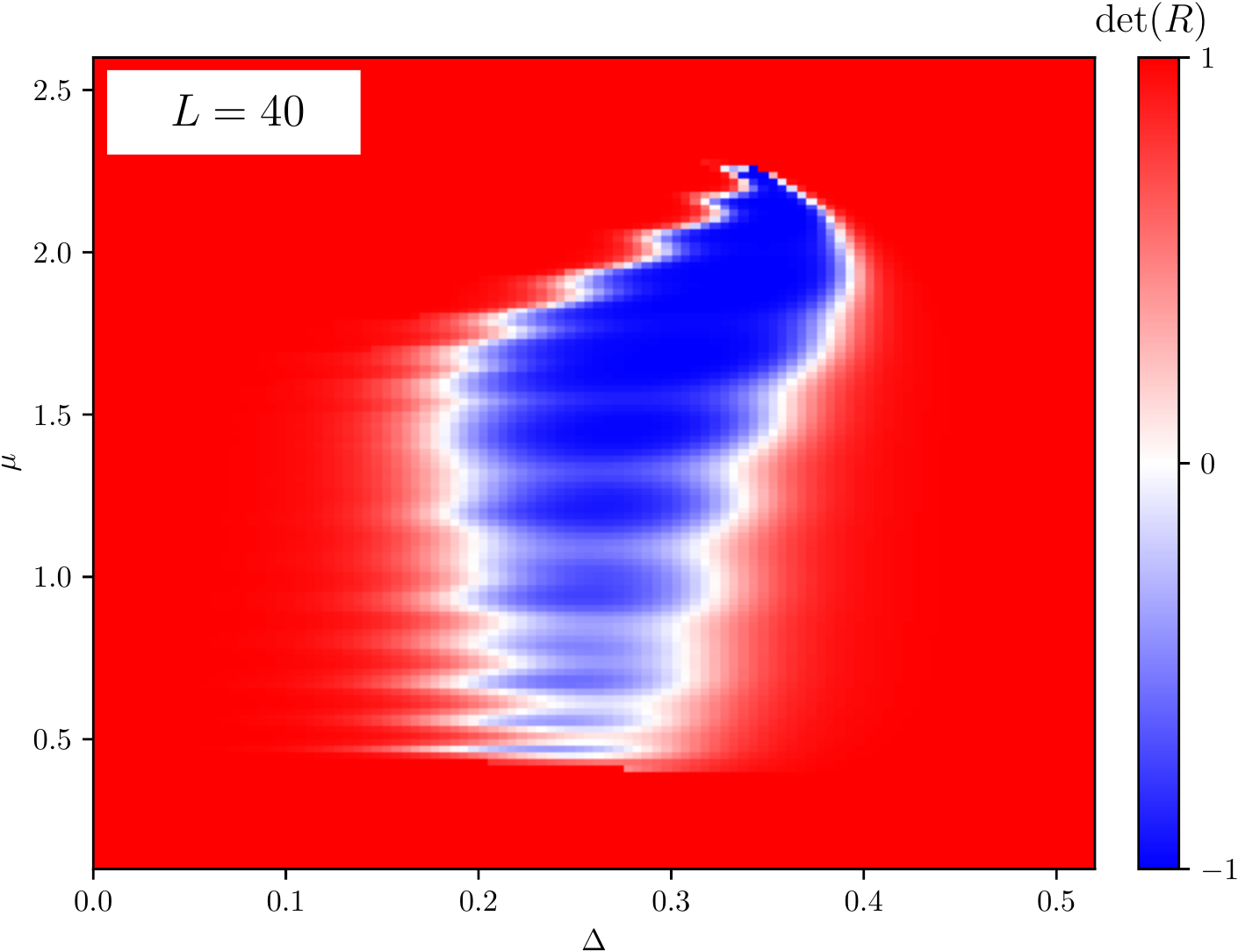}
\includegraphics[width=0.44\textwidth]{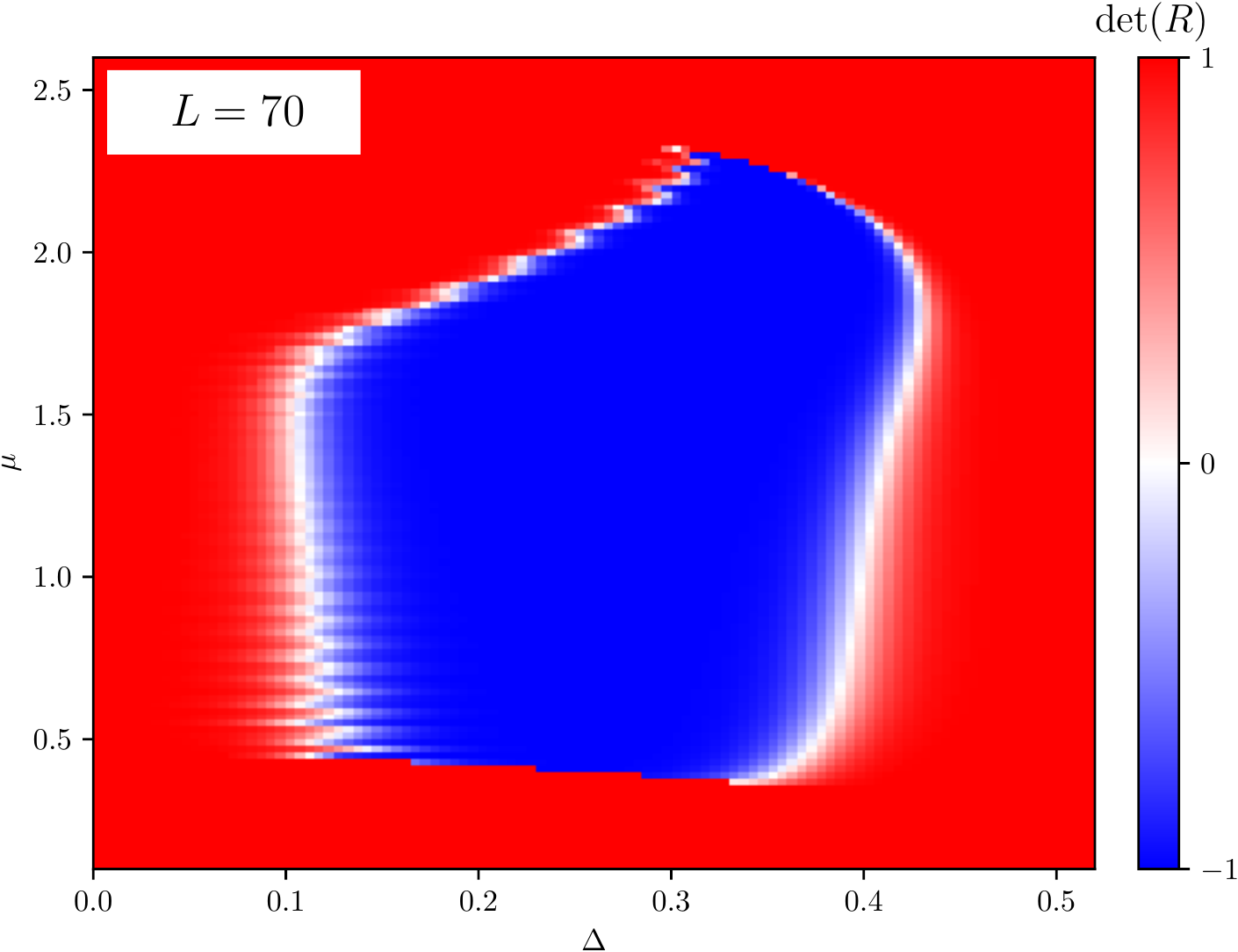}
\includegraphics[width=0.44\textwidth]{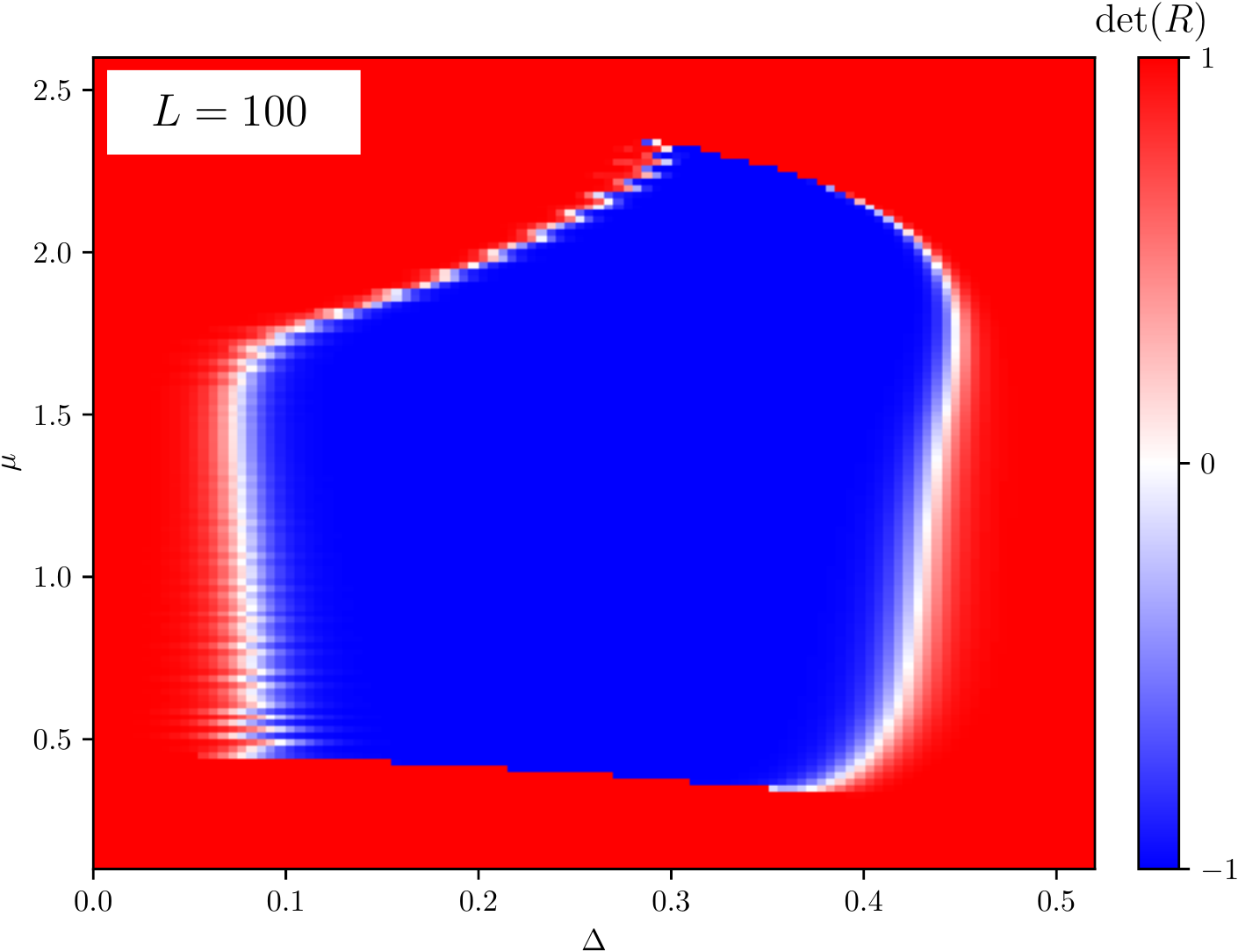}
\includegraphics[width=0.44\textwidth]{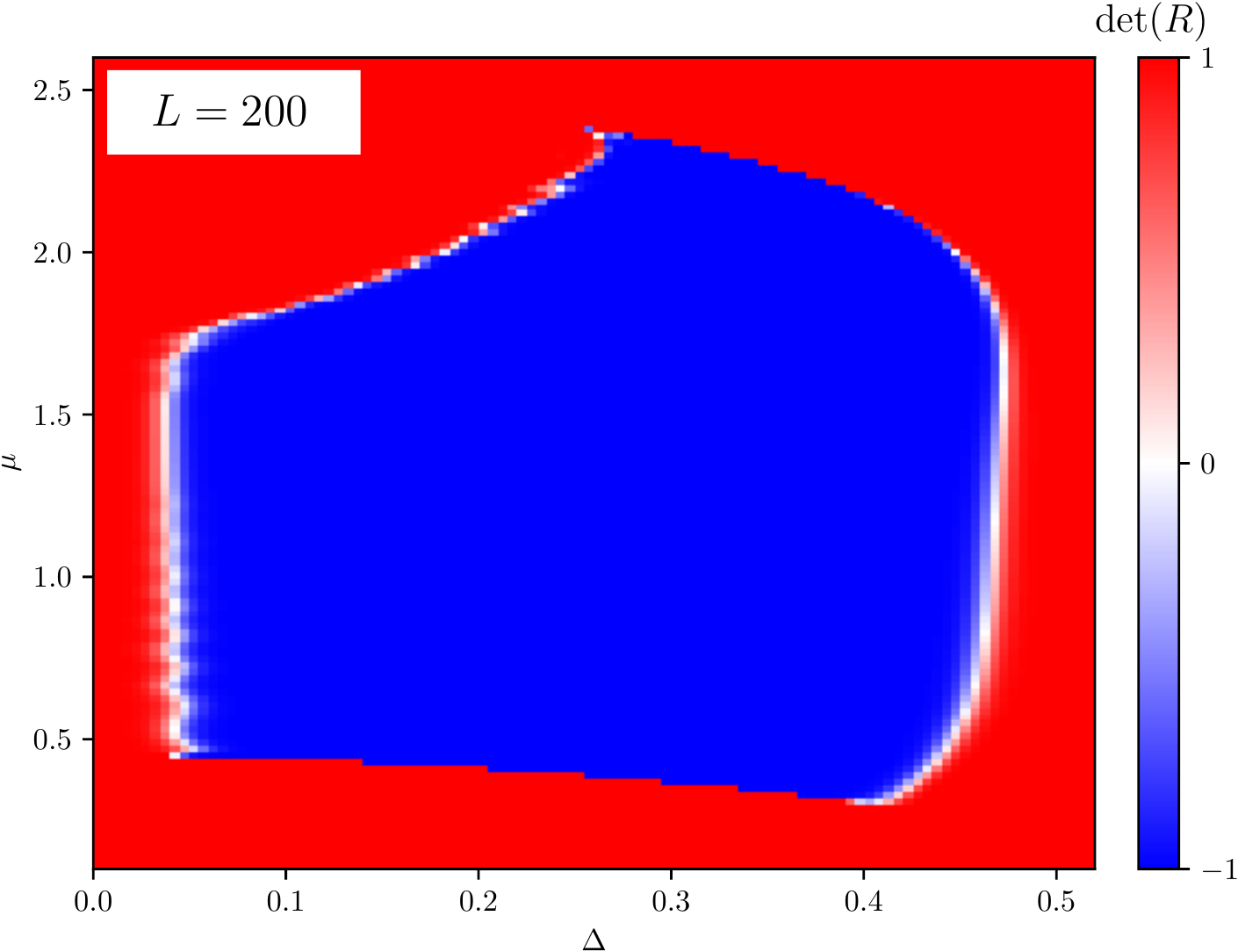}
\caption{$\det(R)$ obtained for $J=1$ and several sizes, ranging from $L=40$ to $L=200$. Computations have been done for the same model parameters as in Fig.~\ref{fig:topo_ph_diags}.
\label{fig:fss}}
\end{figure*}

Pitch vector $q_{*}$ of the ground state (Fig.~\ref{fig:fss_q}) and diagrams of the topological superconducting phase (Fig.~\ref{fig:fss}) clearly indicate that: ({\em i}) $q_{*}$ is hardly affected by nanowire length $L$, ({\em ii}) total area of the topological phase in the parameter space increases with increasing $L$ and ({\em iii}) boundaries of the topological region are much sharper for longer nanowires.
Observations (${i}$) and (${ii}$) suggest that a tendency towards formation of the topological state ({\em topofilia}) should be valid for sufficiently long nanochains. As regards the observation ({\em iii}), it indicates that in the studied system the finite-size effects smooth out the topological transition. This is visible in Fig.~\ref{fig:fss} for $L=40$, where the white area shows such transition between the topologically trivial and nontrivial regions. Finite-size effects are also important for splitting of the Majorana end-modes, when their overlap is sizeable (for short nanowires).

  \input{spiral.bbl}
\end{document}

%% file: spiral.bbl
%